\documentclass[twocolumn,prb,aps]{revtex4}
\usepackage[dvips]{graphicx}
\begin{document}
\title{Analytical continuation of imaginary axis data for optical conductivity }

\pacs{}

\begin{abstract} 
We compare different methods for performing analytical continuation of 
spectral data from the imaginary time or frequency axis to the real frequency
axis for the optical conductivity $\sigma(\omega)$.  We compare the maximum 
entropy (MaxEnt), singular value decomposition (SVD), sampling and Pad\'{e} 
methods for analytical continuation. We also study two direct methods for 
obtaining $\sigma(0)$. For the MaxEnt approach we focus on a recent modification. 
The data are split up in batches, a separate MaxEnt calculation is done for 
each batch and the results are averaged. For the problems studied here, we find 
that typically the SVD, sampling and modified MaxEnt methods give comparable 
accuracy, while the Pad\'{e} approximation is usually less reliable.

\end{abstract}

\author{O. Gunnarsson$^{(1)}$, M. W.  Haverkort$^{(1)}$ and G. Sangiovanni$^{(2)}$}   
\affiliation{
${}^1$Max-Planck-Institut f\"ur Festk\"orperforschung, D-70506 Stuttgart, Germany  \\
${}^2$Institut f\"ur Festk\"orperphysik, Technische Universit\"at Wien, Vienna, Austria
}

\maketitle

\section{Introduction}\label{sec:1}

For strongly correlated systems analytical methods usually involve uncontrolled 
approximations. Therefore stochastical methods such as quantum Monte-Carlo 
(QMC),\cite{Scalapino} quantum cluster methods\cite{DCA} or continuous time 
methods\cite{RubtsovCT} are often used. Apart from statistical errors, such  
methods can produce quite accurate results, but the results are obtained on 
the imaginary axis. A major problem is then the analytically continuing of the 
results to the real axis, which is an ill-posed problem. Small changes in 
the data on the imaginary axis can lead to large changes on the real axis. 
Since the imaginary axis data contain statistical noise, the analytical 
continuation is very difficult.

There are different ways of regularizing this ill-posed problem.
One method combines the Bayesian theory with the maximum entropy 
approach (MaxEnt), which  has been found to be an efficient method 
for analytical continuation.\cite{Jarrell,MEMref}
Other regularizations are used in the singular value decomposition 
(SVD)\cite{SVD,creffield} or stochastic regularization\cite{Rubtsov}
methods. An alternative is provided by making a Pad\'{e} approximation to the data 
as a function of imaginary frequency and then analytically continue the
Pad\'{e} expression to real frequencies.\cite{Vidberg,Baker} A rather different  
approach is to use sampling methods, where a large number of spectra 
are added, weighted by the probability that they correspond to the imaginary
axis data. Such methods have been proposed for $T=0$\cite{Mishchenko} and
finite $T$.\cite{Kiamars} Finally, there are simple approximate methods 
for obtaining the optical conductivity at zero frequency, $\sigma(\omega=0)$
directly from imaginary time or frequency data.

Two-particle correlation functions, such as the dynamical spin or charge 
correlation functions or the optical conductivity, provide important information 
about a variety of properties of the system. These two-particle functions 
are much more difficult to calculate in QMC-like frameworks than the 
one-particle Green's function,\cite{Hanke} and therefore much of the interest 
has focused on the electron Green's function. Here we therefore instead treat 
a two-particle function, the optical conductivity. While we here focus 
on transformation of QMC data from imaginary space to real space, we note 
that there are also QMC methods giving results directly for real frequencies.\cite{Ceperley}

In this paper we compare the Pad\'{e}, SVD, sampling and MaxEnt methods 
for obtaining the optical conductivity from imaginary axis data. 
We define a frequency dependent optical conductivity, $\sigma(\omega)$, 
where $\omega$ is a real frequency. This we refer to as the ``exact'' 
result. This $\sigma(\omega)$ can easily be transformed to the 
imaginary axis, since this is a well-behaved transformation that
can be performed with a high accuracy.  We add statistical noise to the 
data, which then simulate the output of a QMC calculation. The data are 
then transformed  back to the real axis, using the various methods 
for analytical continuation. If the methods work well, we should 
essentially recover the starting $\sigma(\omega)$, the ``exact'' result. 
This way we can judge the accuracy of the different methods. It is important
to compare with a known ``exact'' result, since analytical continuation 
methods can give spurious structures due to noise in the data.
If a certain method A gives more structures than another method B, it is 
hard to judge whether these additional structures are real and method A
is better or they are due to noise and method B is better. This problem
is avoided if ``exact'' results are known. Here we 
construct the ``exact'' $\sigma(\omega)$ using results for the 
two-dimensional Hubbard model as a guide for the general shape.

We find that the SVD, sampling and MaxEnt methods tend to give comparable
accuracy, while the Pad\'{e} approximation often gives worse results.
In particular the Pad\'{e} approximation often overestimates $\sigma(0)$.
One of the direct methods for estimating $\sigma(0)$ (based on Eq.~\ref{eq:2.7} 
in Sec.~\ref{sec:2}) underestimates $\sigma(0)$, in particular for a narrow Drude 
peak, while the other (extrapolating Eq.~(\ref{eq:2.6}) in Sec.~\ref{sec:2} to $\nu=0$)
typically gives better results.
  
In Sec. \ref{sec:2} we present some general results for the optical conductivity.
The different methods for analytical continuation are presented in Sec.
\ref{sec:4} and the results are show in Sec.~\ref{sec:5}.

\section{Optical conductivity and current-current correlation function}\label{sec:2}
The optical conductivity $\sigma(\omega)$ is obtained from the current-current correlation 
function
\begin{equation}\label{eq:2.1}
\Pi(\tau)={1 \over 3N}\langle {\bf j}(\tau)\cdot {\bf j}(0)\rangle,
\end{equation}
where $N$ is the number of sites, ${\bf j}$ is the current operator, ${\bf j}(\tau)
={\rm exp}(H\tau){\bf j}{\rm exp}(-H\tau)$, $\tau$ is imaginary time and 
$\langle ... \rangle$ is the thermodynamic average. We then have   
(setting $\hbar=k_B=1$)
\begin{equation}\label{eq:2.3}
\Pi(\tau)=\int_{-\infty}^{\infty} K(\tau,\omega)\sigma(\omega) d\omega
\end{equation}
where
\begin{equation}\label{eq:2.4}
K(\tau,\omega)={1\over \pi}{\omega e^{-\tau \omega }\over
1-e^{-\beta \omega}},
\end{equation}
is a bosonic kernel 
and $\beta=1/T$.
Alternatively, we can relate $\sigma(\omega)$ to the Fourier transform
$\Pi(\nu)$ of $\Pi(\tau)$
\begin{eqnarray}\label{eq:2.5}
&&\Pi(\nu)\equiv\int_0^{\beta}e^{i \nu \tau}\Pi(\tau)d\tau=-{1\over \pi}\int_{-\infty}^{\infty}
{\omega \over i\nu-\omega}\sigma(\omega)d\omega \nonumber \\
&&= {1\over \pi}\int_{-\infty}^{\infty} {\omega^2 \over \nu^2+\omega^2}\sigma(\omega)d\omega,
\end{eqnarray}
where we have used that $\sigma(\omega)=\sigma(-\omega)$ and 
$\nu=\nu_i=i\nu_0$ is a multiple of $\nu_0=2\pi T$. For large $\nu$ 
we have that $\pi(\nu)\sim \nu^{-2}$. This result
can also be rewritten as
\begin{equation}\label{eq:2.6}
{1\over \pi}\int_{-\infty}^{\infty} 
{\nu \over \omega^2+ \nu^2}\sigma(\omega)d\omega={\Pi(0)-\Pi(\nu)\over \nu}
\equiv \gamma(\nu).
\end{equation}
This provides a convolution of the optical conductivity with a Lorentzian
with the width $\nu$. In particular, if $\sigma(\omega)$ has little variation 
over an energy range of the order of $\nu_0$, $\gamma(\nu=\nu_0)$  provides 
an estimate of $\sigma(0)$. This estimate can be improved by extrapolating 
$\gamma(\nu)$ to $\nu=0$, as discussed below. Alternatively, we can 
use\cite{Kiamars} 
\begin{equation}\label{eq:2.7}
\sigma(0)\approx  {\beta^2\over \pi}\Pi(\tau={\beta \over 2}),
\end{equation}
which is accurate if $\sigma(\omega)$ has little variation over an
energy range of the order of $0.69(2\pi T)$.

\begin{figure}
\vskip-6cm
{\rotatebox{-90}{\resizebox{12.cm}{!}{\includegraphics {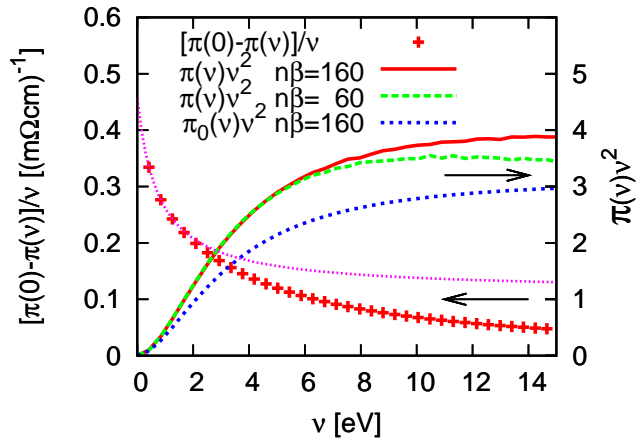}}}}
\caption{\label{fig:1}(color on-line) The current-current correlation 
function $\pi(\nu)$ multiplied by $\nu^2$ as a function of $\nu$ as 
well as the function $\pi_0(\nu)$ calculated without vertex corrections. 
Results are shown for a large number $n\beta=160$ time slices as well as 
for a small $n\beta=60$, illustrating the resulting poor accuracy for large 
$\nu$ in the latter case. The figure also shows $\gamma(\nu)=[\pi(0)-\pi(\nu)]/\nu$ 
[Eq.~(\ref{eq:2.6})] and its extrapolation to $\nu=0$ (thin dotted line), 
which provides an estimate of $\sigma(0)$.
}  
\end{figure}

Typical results for $\Pi(\nu)$ are shown in Fig.~\ref{fig:1} for a
two-dimensional (2d) Hubbard model on a square lattice with nearest ($t=-0.4$ eV)
and second nearest ($t^{'}=0.12$ eV) hopping. The Coulomb interaction is $U=3.2$ eV
and $\beta=15$ eV$^{-1}$. This gives the occupancy 0.95. To obtain a conductivity, 
we have assumed that such 2d 
sheets are stacked on top of each other with a distance $c=6.6$ \AA \ appropriate for 
La$_{2-x}$Sr$_x$CuO$_4$. The data were obtained from a calculation in the 
dynamical cluster approximation (DCA)\cite{DCA} using a cluster with eight sites. 

From Eq.~(\ref{eq:2.5}) we can see that $\Pi(\nu)\sim \nu^{-2}$ for large
$\nu$. In Fig.~\ref{fig:1} we show $\Pi(\nu)\nu^2$, which indeed saturates 
for large $\nu$.  From Eq.~(\ref{eq:2.5}) we expect this to happen when $\nu$ 
is much larger than a typical energy scale of $\sigma(\omega)$, which in this 
case has peaks at $\omega=0$ and $\omega\approx \pm U=\pm 3.2$ eV. In agreement with this, 
Fig.\ref{fig:1} shows saturation for $\nu$ of the order of 10  eV. 
For larger values of $\nu$, $\Pi(\nu)$ essentially just gives information
about $\int \omega^2 \sigma(\omega)d \omega$. The figure also shows results 
for $\Pi_0(\nu)$, which is calculated neglecting all vertex corrections.
$\Pi_0$ is then obtained simply as a product (bubble) of two (dressed) Green's 
functions. Even for large $\nu$, $\Pi$ and $\Pi_0$ are different. This can be
understood from Eq.~(\ref{eq:2.5}). Although both behave as $\nu^{-2}$,
the prefactor is different.

Fig.~\ref{fig:1} also shows $[\Pi(0)-\Pi(\nu)]/\nu$ [Eq.~(\ref{eq:2.6})],
providing an estimate of $\sigma(0)$. To improve this estimate we extrapolate
to $\nu=0$. For small values of $\nu$, $\Pi(\nu)$ depends mainly on $\sigma(w)$
 for small $\omega$.  We allow for the possibility that $\sigma(\omega)$ has 
a Drude like peak at $\omega=0$ by using the Ansatz
\begin{equation}\label{eq:a1}
\sigma(\omega)=a+b{\Gamma/\pi\over \omega^2+\Gamma^2},
\end{equation}
where we have also added a constant $a$. In Fig.~\ref{fig:1} we have fitted this
form to the results for the lowest three non zero values of $\nu$. This extrapolation
greatly improves the estimate, as can be seen from the examples below.

\section{Methods}\label{sec:4}

\subsection{Pad\'{e} Approximation}\label{sec:4a}

In the Pad\'{e} approximation a function $f(z)$ in the complex plane, $z$, is 
described as the ratio between to polynomials $P(z)$ and $Q(z)$, $f(z)=P(z)/Q(z)$. 
The function is fitted to the output of a QMC calculation so that the 
results for certain imaginary frequencies $\nu_n$ are reproduced exactly.
The analytical continuation is then performed by evaluating the function
on the real axis. In the context of Green's functions this has in particular 
been used by Vidberg and Serene.\cite{Vidberg} They fit to $N$ data points,
using a construction which for an even $N$ leads to a polynomial
$Q$ which is one order higher than $P$, so that $P/Q$ behaves as $1/z$
for large $z$. This is appropriate for Green's functions, considered by 
them, but not necessarily for the response functions considered here, 
which behave as $1/z^2$ for large $z$. We have therefore constructed 
a Pad\'{e} approximation where $Q$ is two orders higher than $P$, which is 
used in the following. This construction requires $N$ to be odd. 
For the special case considered by Vidberg and Serene there are simple 
formulas for generating the polynomials,\cite{Vidberg} while this is 
somewhat more complicated in the general case.\cite{Baker}

In fitting the $P$ and $Q$ to $N$ data points, we have used data for 
one negative frequency,$-\nu_0\equiv -2\pi T$, and the $N-1$ lowest nonnegative
frequencies. This typically gives more stable results than using only
nonnegative frequencies. On the other hand, using positive and negative 
frequencies symmetrically tends to put poles close to the real axis and
gives poor spectra on the real axis. One negative frequency therefore 
often appears to be a good compromise.

\subsection{Singular value decomposition}\label{sec:4d}

A widely used technique for inverse problems is the singular
value decomposition (SVD).\cite{SVD,creffield} Here we essentially 
follow Creffield {\it et al.},\cite{creffield} except that we work 
in imaginary frequency space rather than in imaginary time space,
for reasons discussed Sec.~\ref{sec:5}.  In the SVD method, the
real frequency space is spanned by a set of eigenvectors. The kernel
in Eq.~(\ref{eq:2.5}) is discretized, giving
\begin{equation}\label{eq:3.11}
\Pi(\nu_i)=\sum_{j=1}^{N_{\omega}}K_{ij}\sigma(\omega_j), \hskip1cm
i=1,N_{\nu }.
\end{equation}
If the data for different imaginary frequencies $\nu_i$ are uncorrelated,
as is the case here, we introduce the eigenfunctions of the operator 
$KK^{\dag}$
\begin{equation}\label{eq:3.12}
\sum_{j=1}^{N_{\omega}} \sum_{l=1}^{N_{\nu }}K_{ij}K^{*}_{lj}v^k_l
=\alpha_k^2 v^k_i, \hskip1cm i=1,N_{\nu}\\ 
\end{equation}
We introduce vectors $u_k$ 
\begin{equation}\label{eq:3.13}
K^{\dag}v^k=\alpha_ku^k,
\end{equation}
which satisfy
\begin{equation}\label{eq:3.13a}
Ku^k=\alpha_k v^k.
\end{equation}
The spectral function can then be expanded as
\begin{equation}\label{eq:3.14}
\sigma(\omega_j)=\sum_{k=1}^{N_{\nu }}{1\over \alpha_k}u^k_j
\sum_{i=1}^{N_{\nu }}(v^k_i)^{\ast}\Pi(\nu_i)
\end{equation}
This expansion is very ill-behaved, since some of the eigenvalues 
are very small. The expansion is therefore truncated so that only eigenvalues
are considered for which 
\begin{equation}\label{eq:3.14a}
\alpha^k/\alpha_1> \sigma_0,
\end{equation}
where $\alpha_1$ is the largest eigenvalue and $\sigma_0$ is the accuracy 
of the data. In this way we only consider the $n_{\nu}$ eigenvectors with 
the largest eigenvalues. To further improve the method, the kernel $K$ is 
multiplied by a ``support'' function, which is equal to one in the range 
where $\sigma(\omega)$ is expected to be large and vanishes smoothly 
outside this region. Here we have used the function 
$1/(1+(\omega/\omega_0)^8)$, where $\omega_0=5$ was used.

\subsection{Maximum Entropy}\label{sec:4b}

A popular method for analytical continuation is the maximum entropy
method (MaxEnt).\cite{Jarrell} This method is based on Bayes's theorem\cite{Bayes}
\begin{equation}\label{eq:3.1}
{\rm Pr}[\sigma,\Pi]={\rm Pr}[\sigma|\Pi]{\rm Pr}[\Pi]={\rm Pr}
[\Pi |\sigma]{\rm Pr}[\sigma],
\end{equation}
where Pr$[\sigma,\Pi]$ is the joint probability that the spectral function
is $\sigma(\omega)$ and that the QMC calculation gives the correlation
function $\Pi(\nu)$. While the MaxEnt method usually is formulated
for imaginary time $\tau$, we here formulate it for imaginary frequency $\nu$,
for reasons discussed in Sec.\ref{sec:5}. Pr $[\sigma|\Pi]$ is the 
conditional probability that the spectral function is $\sigma(\omega)$ 
provided that the correlation function $\Pi(\nu)$ was obtained from the 
QMC calculation. From this one obtains\cite{Skilling}
\begin{equation}\label{eq:3.2}
{\rm Pr}[\sigma|\Pi]={{\rm Pr} [\Pi|\sigma]{\rm Pr}[\sigma]
\over {\rm Pr}[\Pi]}.
\end{equation}
This rewriting converts the ill-posed problem of determining
$\sigma(\omega)$ given $\Pi(\nu )$ into the much easier problem
of determining $\Pi(\nu )$ given $\sigma(\omega)$. Pr$[\Pi]$
is a normalization factor, which is independent of $\sigma(\omega)$, 
and therefore is no complication. The remaining issue is then how to choose 
Pr$[\sigma]$, which represents our prior knowledge about $\sigma(\omega)$. 
If we put this probability to a constant and then maximize the liklihood 
function Pr$[\Pi|\sigma]$ the result is typically very bad, resulting in a 
saw-tooth type of spectra.\cite{Skilling} In MaxEnt one therefore defines   
the prior probability in terms of a maximum entropy function
\begin{equation}\label{eq:3.3}
S=\int d\omega \lbrace \sigma(\omega)-m(\omega)-\sigma(\omega){\rm ln} 
 { \sigma(\omega) \over m(\omega) } \rbrace,
\end{equation}
where $m(\omega)$ is a default model. Other definitions are also 
possible.\cite{White} In the MaxEnt method the quantity
\begin{equation}\label{eq:3.4}
{\rm Pr} [\Pi|\sigma]e^{\alpha S}
\end{equation}
is maximized, using an appropriate value for $\alpha$.\cite{Jarrell}
Here we have chosen $\alpha$ according to the classic MaxEnt method, 
using  a flat prior for $\alpha$.\cite{Jarrell} 

We sometimes find that this approach leads to unphysical oscillations 
in $\sigma(\omega)$. We have shown that the reason is that the MaxEnt
method sometimes chooses an $\alpha$ which attaches too much significance
to the noise in the data. This problem can be avoided by using a modification
of the MaxEnt method.\cite{batching} We split the data for $\Pi(\nu)$ in 
several batches and perform a MaxEnt calculation for each batch. 
These results are then averaged. Typically, but not always, this leads to 
better results than averaging the data sets and then performing just 
one MaxEnt calculation for the average.\cite{batching}

The choice of default model can influence the outcome substantially.
Here we have chosen a ``reasonable'' but structureless model (see
Sec. \ref{sec:5}).  Using a model more similar to the actual spectrum 
improves the result. If the spectrum is calculated for several $T$, 
the result for a higher $T$ can be used as the default model for a 
lower $T$. This can improve the results without introducing undue bias. 
Since we only consider one $T$ here, we have not followed that approach.

\subsection{Sampling method}\label{sec:4c}

The MaxEnt method avoids the saw-tooth problem, but the definition of entropy
requires the introduction of a default model, which can bias the output. 
Instead we can average\cite{Skilling} over
Pr$[\sigma|\Pi]$
\begin{equation}\label{eq:m3}
\langle \sigma \rangle=\int \sigma {\rm Pr} [\sigma|\Pi] {\scriptstyle D} \sigma,
\end{equation}
where ${\scriptstyle D }\sigma$ indicates a functional integral over all
$\sigma(\omega)$. ${\rm Pr} [\sigma|\Pi]$ is given by Eq.~(\ref{eq:3.2}),
where we furthermore put Pr$[\sigma]\equiv$ constant for all nonnegative 
$\sigma(\omega)$. Thus we assume that this is our only prior knowledge 
of $\sigma(\omega)$. In Ref.~\onlinecite{Kiamars} we worked in
imaginary time space. For reasons discussed in sec.~\ref{sec:5} we here work
in imaginary frequency space. Then the likelihood function is given 
by\cite{Skilling}
\begin{eqnarray}\label{eq:m4}
&&{\rm Pr}[\Pi|\sigma]={1\over \Pi_{i=1}^{n_{\nu}}(2\pi\tilde \sigma_i)} \\
&&\times {\rm exp} \lbrace-\sum_{i=1}^{n_{\nu}} [\Pi(\nu_i)-\Pi_{\sigma}(\nu_i)]^2/
(2\tilde \sigma_i^2)\rbrace,
\nonumber
\end{eqnarray}
where $\tilde \sigma_i$ is the accuracy of the data $\Pi(\nu_i)$, and
$\Pi_{\sigma}(\nu_i)$ is the transformation of $\sigma(\omega)$ to 
imaginary frequencies.

\begin{figure}
{\rotatebox{-90}{\resizebox{5.7cm}{!}{\includegraphics {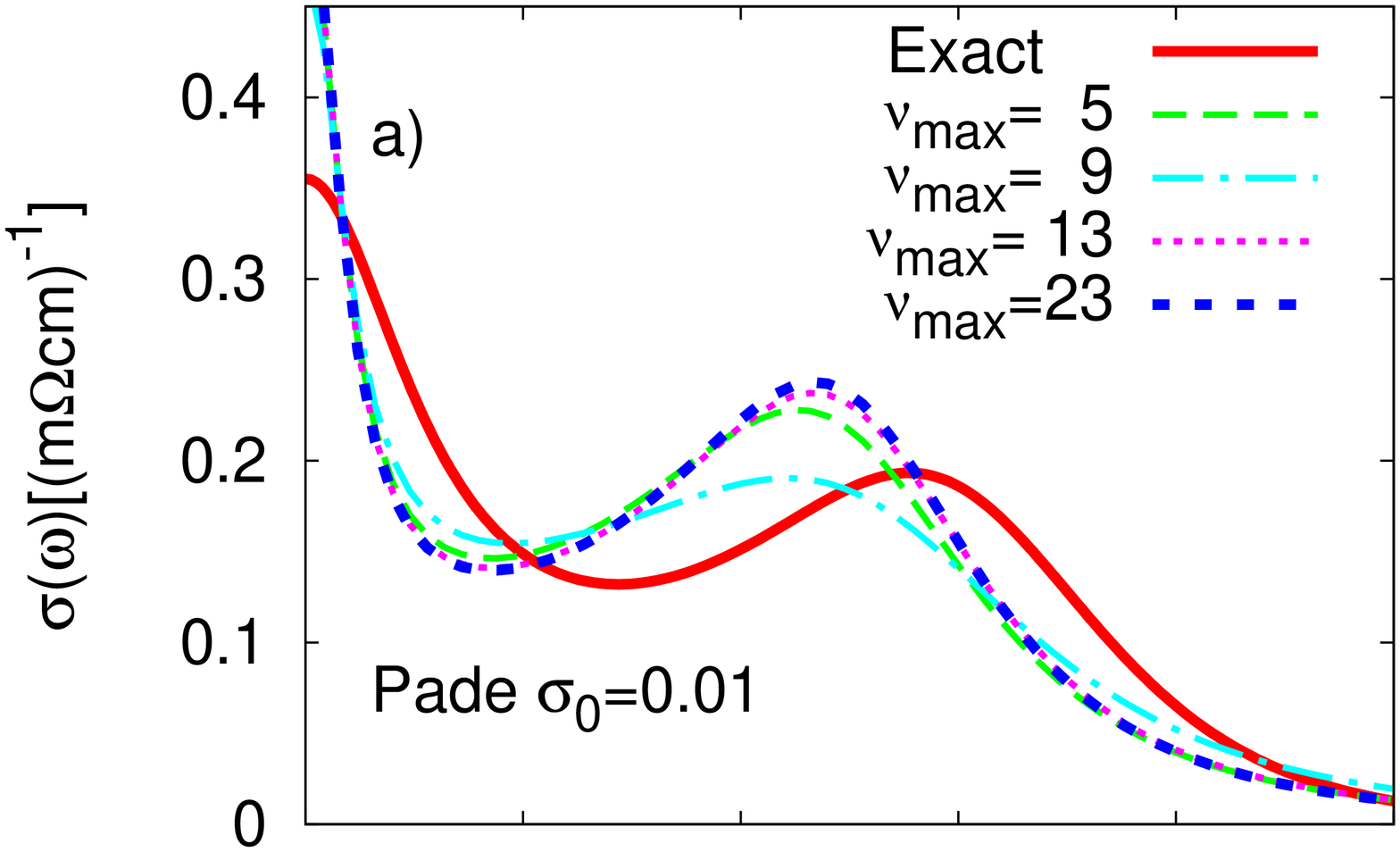}}}}
\vskip-1.34cm
{\rotatebox{-90}{\resizebox{5.7cm}{!}{\includegraphics {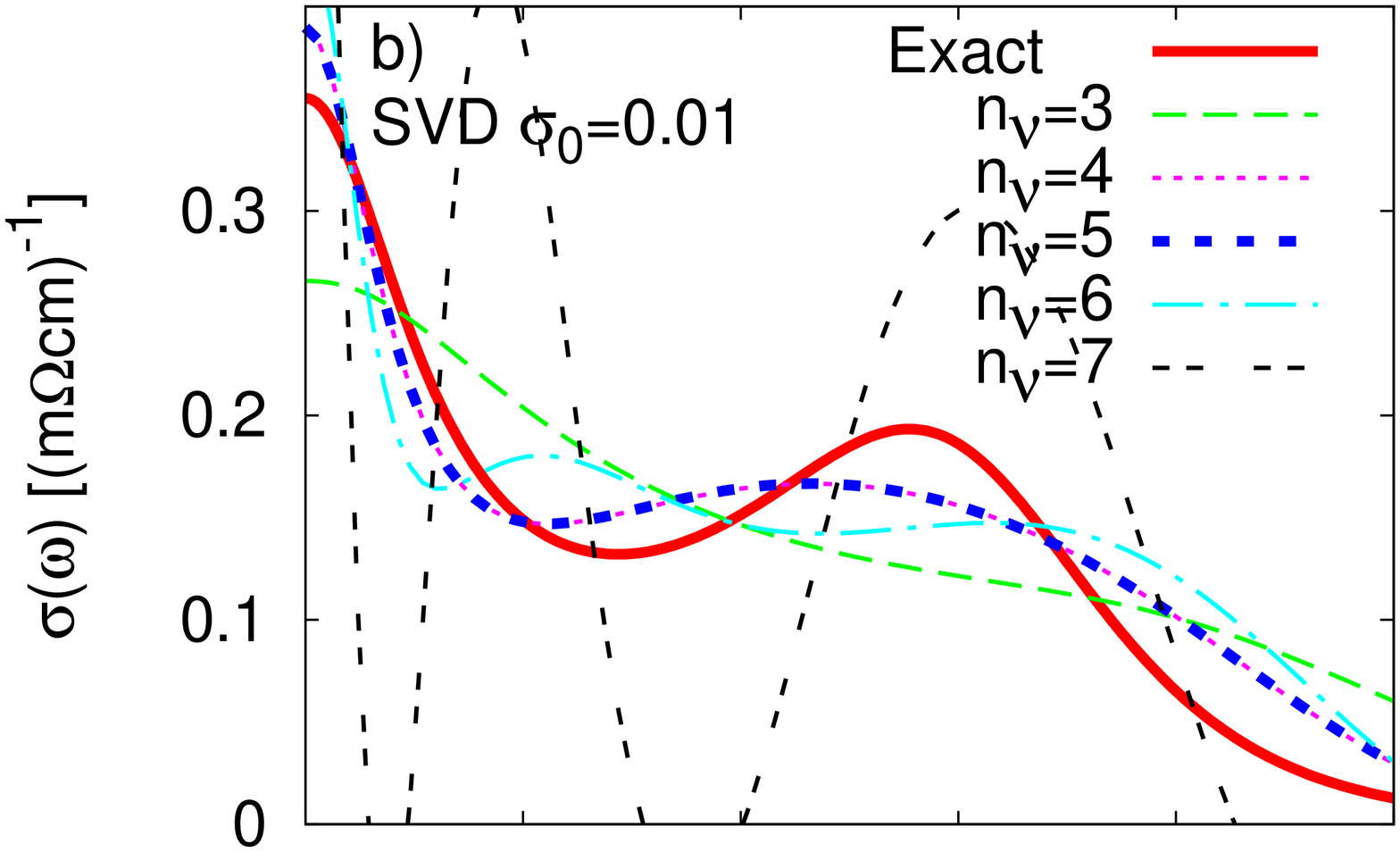}}}}
\vskip-1.34cm
{\rotatebox{-90}{\resizebox{5.7cm}{!}{\includegraphics {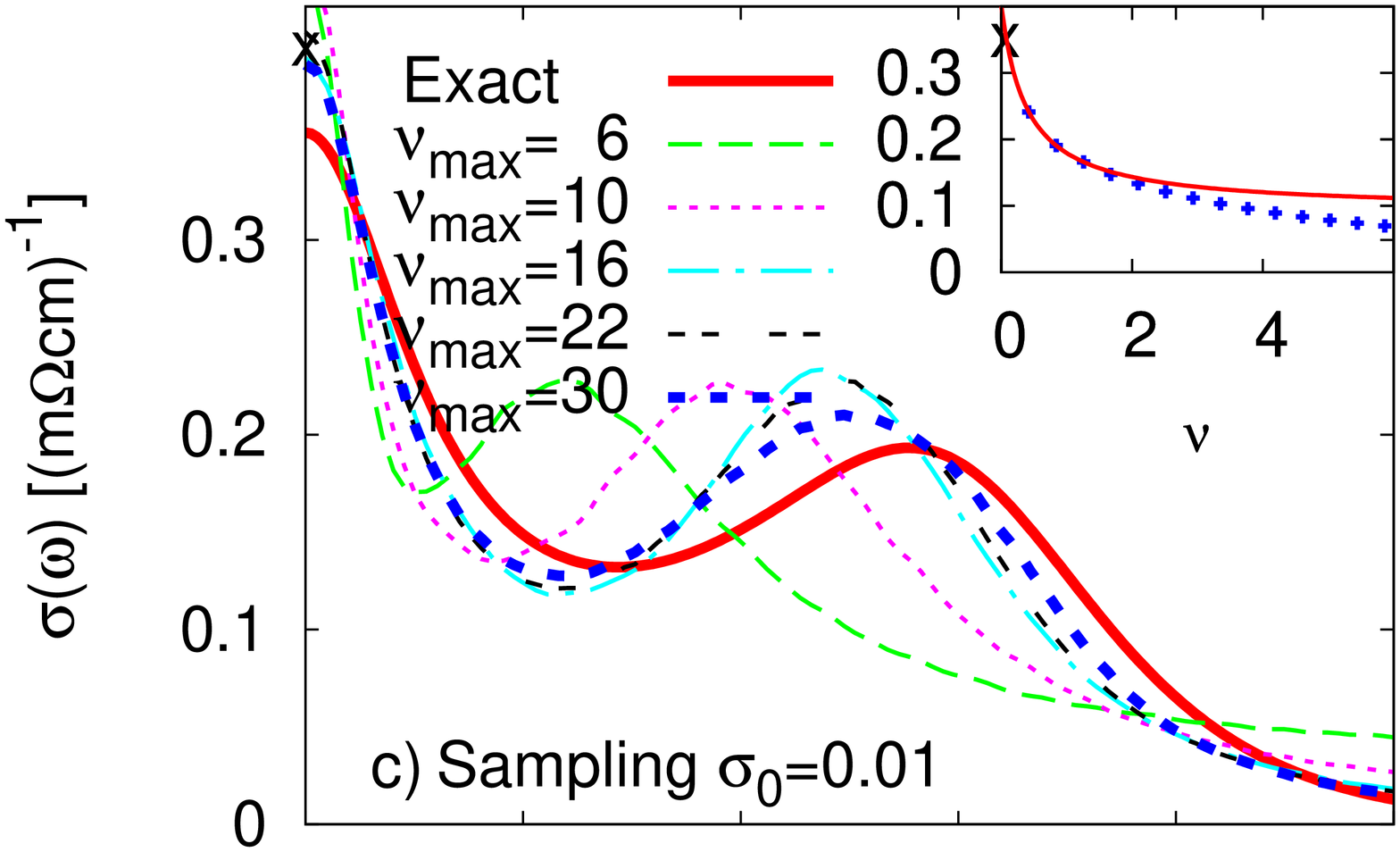}}}}
\vskip-0.78cm
{\rotatebox{-90}{\resizebox{5.70cm}{!}{\includegraphics {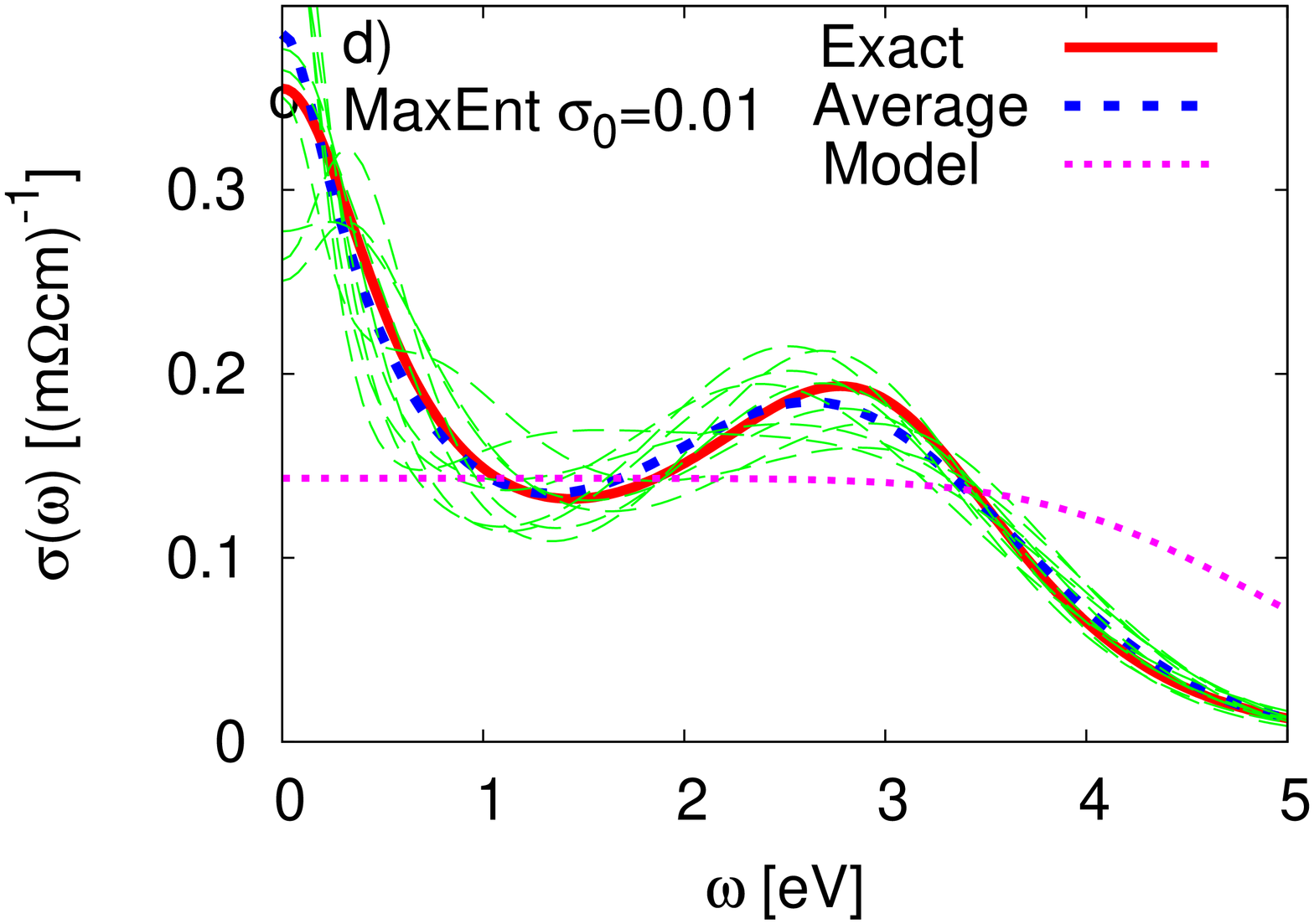}}}}
\caption{\label{fig:2}
The optical conductivity for model (\ref{eq:4.1}) using $\Gamma_1=0.60$
and $\sigma_0=0.01$
according to the Pad\'{e} (a), the SVD method (b), the
sampling (c) and the MaxEnt (d) methods compared with the exact results. 
Figs. a)  and c) show results for different values of the maximum 
frequency $\nu_{\rm max}$ considered and Fig. b) for different values 
of $n_{\nu}$.  Fig. d) shows results both for each individual sample 
(thin lines) and the average over all 10 samples as well as the model used.
The thick line in (b) indicates the optimum value of $n_{\nu}=5$ and the
thick line in (c) the largest value of $\nu_{\rm max}=30$ considered here.
The $x$ in the main part of figure c) shows the estimate of $\sigma(0)$ by 
extrapolating $[\Pi(0)-\Pi(\nu)]/\nu$ to zero. 
This is  illustrated by the inset in figure c), where the symbol $\times$ gives the 
exact value of $\sigma(0)$. The symbol $o$ in (d) is the estimate of $\sigma(0)$  
based on $\Pi(\beta/2)$ [Eq. (\ref{eq:2.7})]. 
}  
\end{figure}

\section{Results}\label{sec:5}

To study methods of analytical continuation, we choose a model of 
$\sigma(\omega)$ on the real frequency axis, using calculations for
a two-dimensional (2d) Hubbard model as a guide. Using
Eq.~(\ref{eq:2.5}), the corresponding $\Pi(\nu)$ can easily be 
calculated. This is a well-behaved and stable transformation.  
We generate results for the 60 smallest nonnegative frequencies.
We add random noise to this calculated $\Pi(\nu)$,
\begin{equation}\label{eq:2.8}
\Pi_{\mu}(\nu_i)=\Pi(\nu_i)(1+r_{\mu,i}),
\end{equation}
where $r_{\mu,i}$ has a Gaussian distribution with the width $\sigma_0$.
This simulates the data that may be obtained from a QMC calculation
by solving the Bethe-Salpeter equation. We generate 10 different sets of 
data using different random numbers for each set. 

In the DCA approach, $\Pi(\nu)$ is Fourier transformed to obtain 
$\Pi(\tau)$. This may require knowledge of $\Pi(\nu)$ for frequencies 
where the calculation is not very accurate. Although this problem can 
usually be circumvented by using the asymptotic behavior of $\Pi(\nu)$
for large $\nu$, it then seems easier to work directly in $\nu$-space. 
Then if necessary, we can then decide to use fewer values of $\nu$ than 
is needed to converge the Fourier transform and only use values which we 
believe are accurate. Specifically for the present calculation, a Fourier 
transform to $\tau$-space would lead to additional complications. Although 
by construction the present $\Pi_{\mu}(\nu_i)$ has a perfectly Gaussian 
noise which is uncorrelated for different values of $\nu_i$, the Fourier 
transformed data would have correlation between different $\tau$-points. 
Methods working in $\tau$-space and methods working in $\nu$-space would then 
have data of different quality.  To be able to compare all methods 
on an equal footing, we have therefore formulated them in imaginary 
frequency space, which essentially involves using kernels appropriate
for this space. 

The $\Pi_{\mu}(\nu_i)$ data are then analytically 
continued back to the real axis, using methods of interest. Since we know 
the exact result, namely the $\sigma(\omega)$ we started from, we can test 
the accuracy of the methods. 

For the MaxEnt method we analytically continued
each data set and then took the average.\cite{batching} As discussed above,
the reason is that the MaxEnt method tends to attach too much significance
to the noise. The batching method reduces the importance of the noise at
the cost of using data with a lower accuracy. For the SVD (with the condition
in Eq.~[\ref{eq:3.14a}]) and sampling methods we have not noticed any tendency 
to overemphasizing the noise. Therefore we averaged the data before doing
the analytical continuation to get data with the highest possible accuracy.
For the Pade method with many data points on the imaginary axis there is a 
strong tendency to overemphasize the noise. However, we have not noticed 
any general improvement by "batching" the data, and therefore also for the 
Pade approximation we  averaged the data before doing the analytical continuation.

The optical conductivity $\sigma(\omega)$ typically has peaks at $\omega=0$ 
and  at approximately $\omega=\pm U$, where $U$ is the Hubbard on-site 
Coulomb interaction. We therefore use the real axis $\sigma(\omega)$
\begin{eqnarray}\label{eq:4.1}
&&\sigma(\omega) 
=\lbrace {W_1  \over 1+ (\omega /\Gamma_1)^2} 
+{W_2  \over 1+[(\omega-\epsilon)/ \Gamma_2]^2} \\
&&+{W_2  \over 1+[(\omega+\epsilon)/ \Gamma_2]^2}\rbrace
{1\over 1+(\omega / \Gamma_3)^6 } \nonumber 
\end{eqnarray}
Here $\Gamma_3\gg (\Gamma_1, \Gamma_2$) cuts off $\sigma(\omega)$ for
large $\omega$. Otherwise $\Pi(\nu)$ would not decay as $\nu^{-2}$,
as it should. Here we let $\omega$ and $\Gamma_i$ have the unit eV
and $\sigma$ the unit $({\rm m}\Omega {\rm cm})^{-1}$. 
Since the smallest nonzero frequency $\nu=\nu_0 
\equiv 2\pi T$, we expect structures on an energy scale much 
smaller than $\nu_0$ to be described very poorly. Here we use 
$T=1/15$, giving $\nu_0=0.42$. We then choose two different 
models with $\Gamma_1=0.30$ and 0.6, respectively. For both models
we use $\Gamma_2=1.2$, $\Gamma_3=4$ and $\epsilon=3$. We use the 
weights $W_1=0.3$ and $W_2=0.2$.

\begin{figure}
{\rotatebox{-90}{\resizebox{5.7cm}{!}{\includegraphics {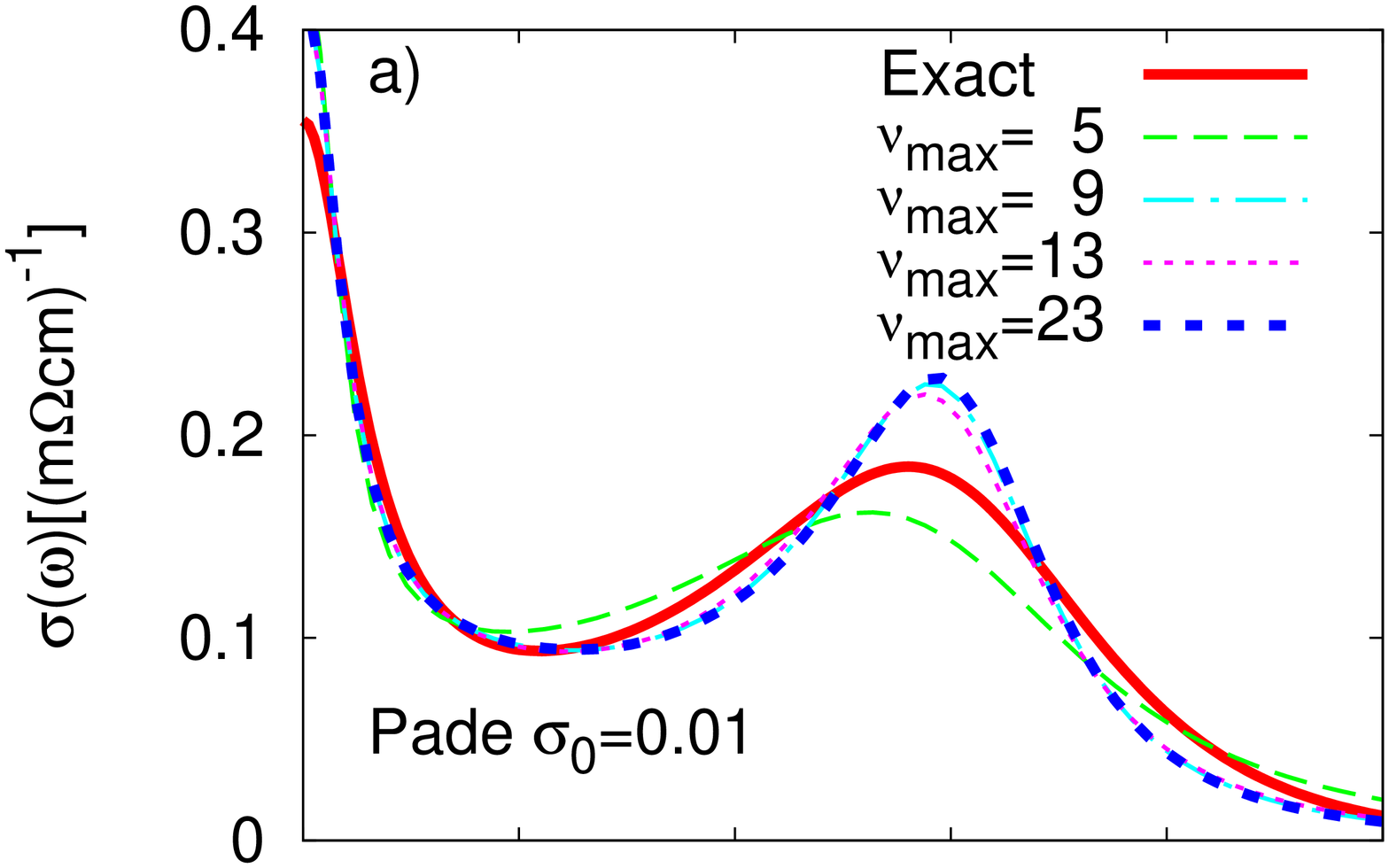}}}}
\vskip-1.34cm
{\rotatebox{-90}{\resizebox{5.7cm}{!}{\includegraphics {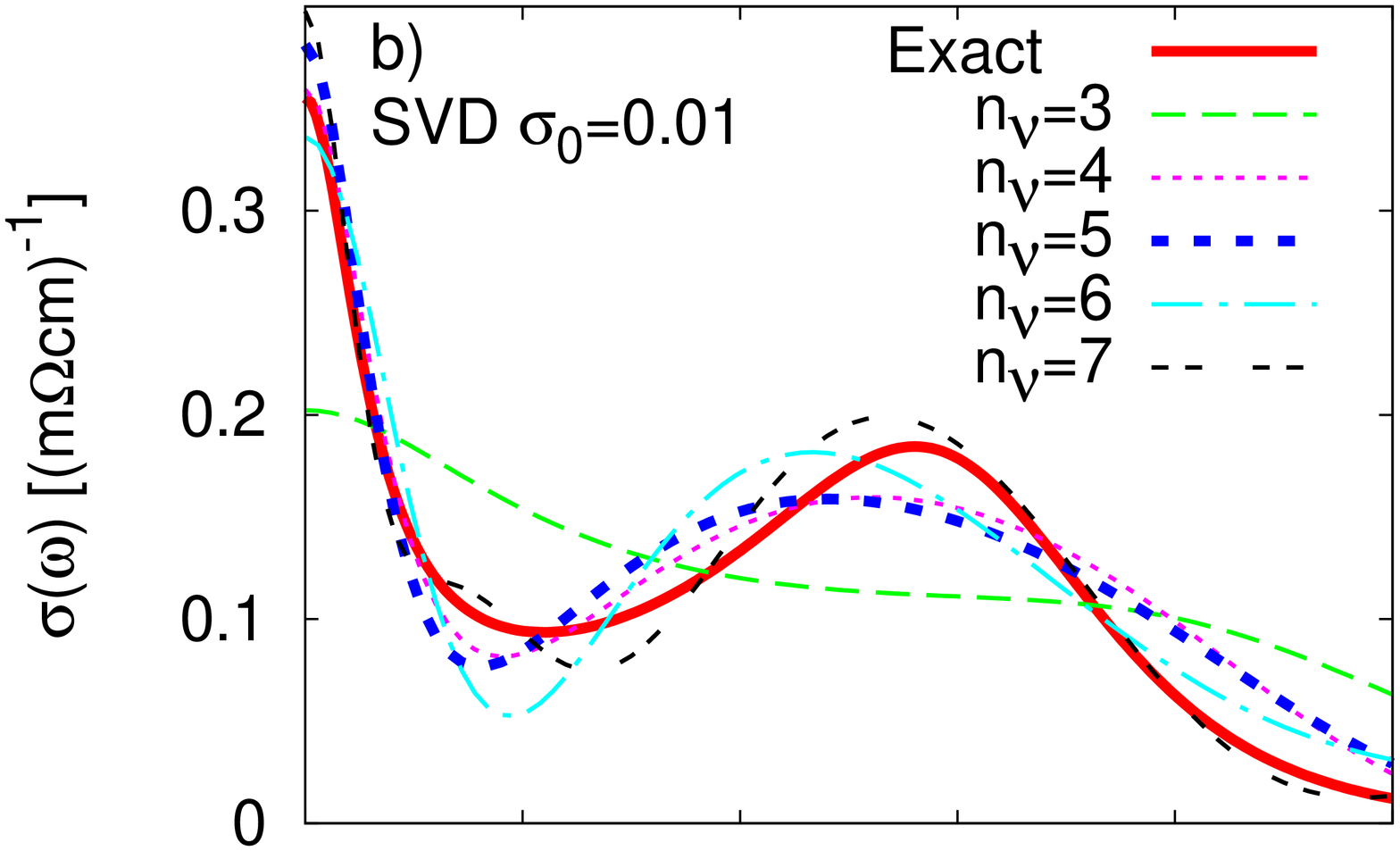}}}}
\vskip-1.34cm
{\rotatebox{-90}{\resizebox{5.7cm}{!}{\includegraphics {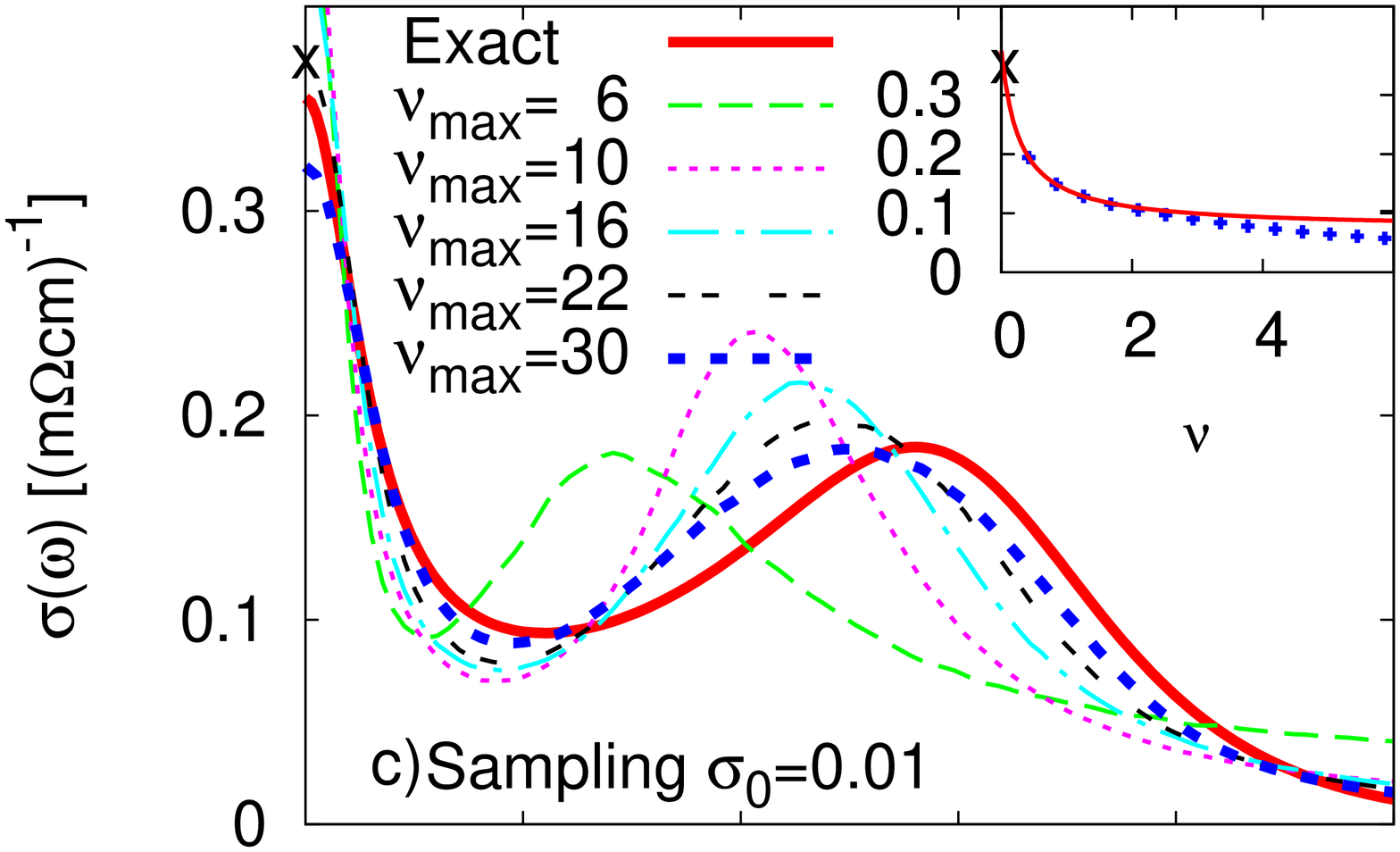}}}}
\vskip-0.78cm
{\rotatebox{-90}{\resizebox{5.70cm}{!}{\includegraphics {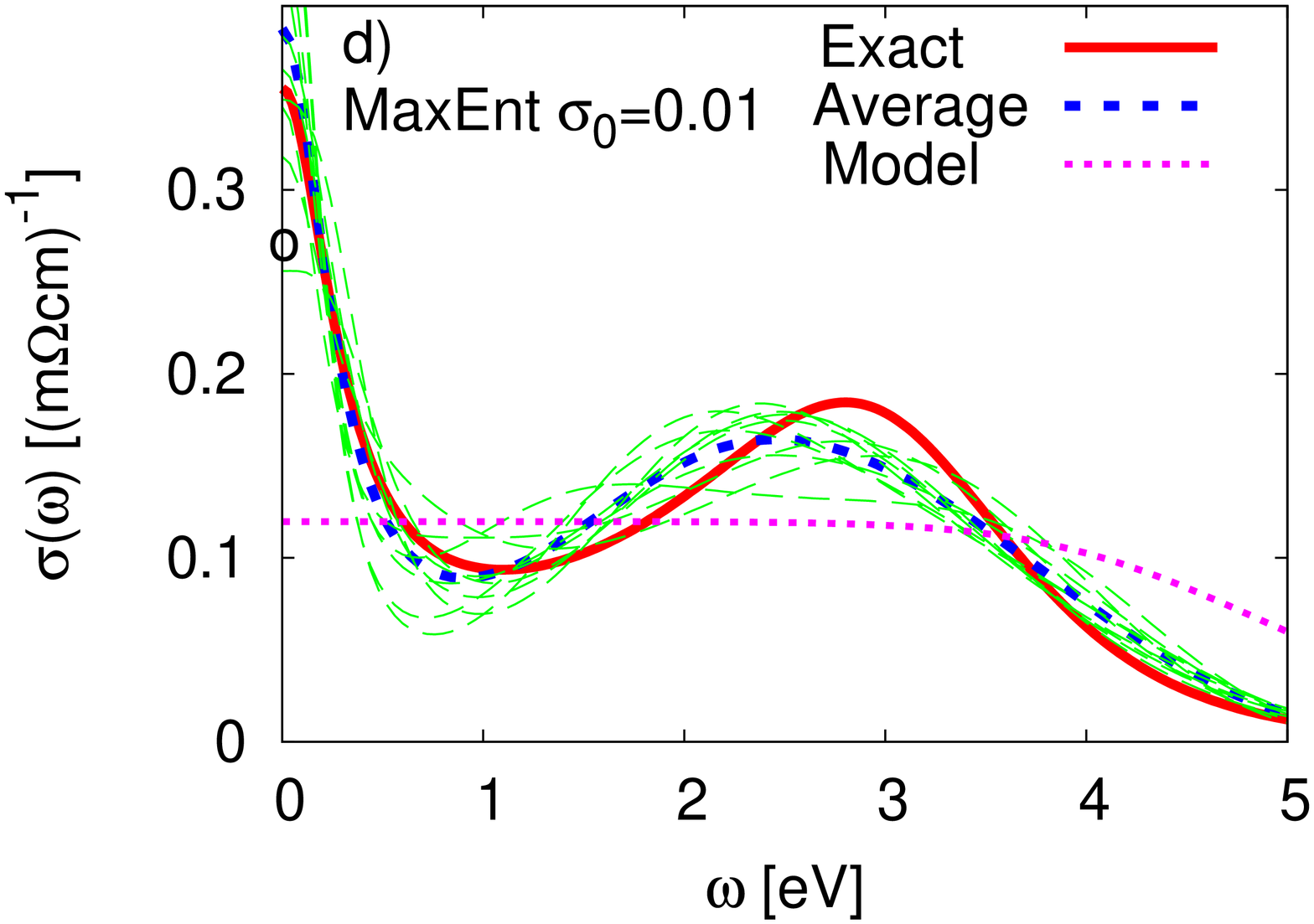}}}}
\caption{\label{fig:3}
The same as Fig. \ref{fig:2} but for $\Gamma_1=0.3$ ($\sigma_0=0.01$). 
}  
\end{figure}

Fig.~\ref{fig:2} shows results for $\Gamma_1=0.6$ and data with
relatively good accuracy $\sigma_0=0.01$. Fig.~\ref{fig:2}a shows
results according to the Pad\'{e} approximation for different numbers
($\nu_{\rm max}$) of frequencies. For $\nu_{\rm max}=5$  
the spectrum is rather structureless and the peak at $\omega=3$ is
not well described. Since $\nu_{\rm max}=5$ corresponds to an
imaginary frequency 1.7, smaller than the energy scale for
the structures on the real axis, this is not surprising.
As $\nu_{\rm max}$ is increased and more information is added, 
this peak is formed, although at too small energy. The  
peak at $\omega=0$ is also not very well described.  

Fig.~\ref{fig:2}b shows results according to the SVD method. The 
eigenvalues $\alpha_k$ in Eq.~(\ref{eq:3.12}) are in this case 0.50, 0.12, 
0.048, 0.020, 0.0067, 0.0020, 0.00055, 0.00015...  The optimal value of 
$n_{\nu}$ according to the criterion in Sec. \ref{sec:4d} and $\sigma_0=0.01$ 
is then 5 and the corresponding results are  shown by the thick line.
Results are also shown by thin lines for $n_{\nu}$ = 3, 4, 6 and 7.
$n_{\nu}=3$ is too small, and misses most of the structures. The values
$n_{\nu}=4$ gives similar results as $n_{\nu}=5$, while $n_{\nu}=6$
gives some unphysical oscillations and $n_{\nu}=7$
puts in large spurious structure, giving too much weight to the noise. 

Fig.~\ref{fig:2}c shows results from the sampling method. For small values 
of $\nu_{\rm max}$ the structures are poorly described, and, in particular,
the Hubbard peak is placed at a too low energy. As discussed for the Pad\'{e} 
approximation, this is not surprising since only information for small
imaginary frequencies is used. As $\nu_{\rm max}$ is increased the 
description improves. For a large $\nu_{\rm max}=30$, shown  by the 
thick line, the description is rather good. The inset shows the quantity
$\gamma(\nu)$ in Eq.~(\ref{eq:2.6}) and the extrapolation to $\nu=0$,
giving an estimate of $\sigma(0)$. The symbol $\times$ in the inset gives the exact 
result and the $\times$ in the main figure \ref{fig:2}c shows the result
estimated from this extrapolation. This estimate is in this case somewhat
too large. 

\begin{figure}
{\rotatebox{-90}{\resizebox{5.7cm}{!}{\includegraphics {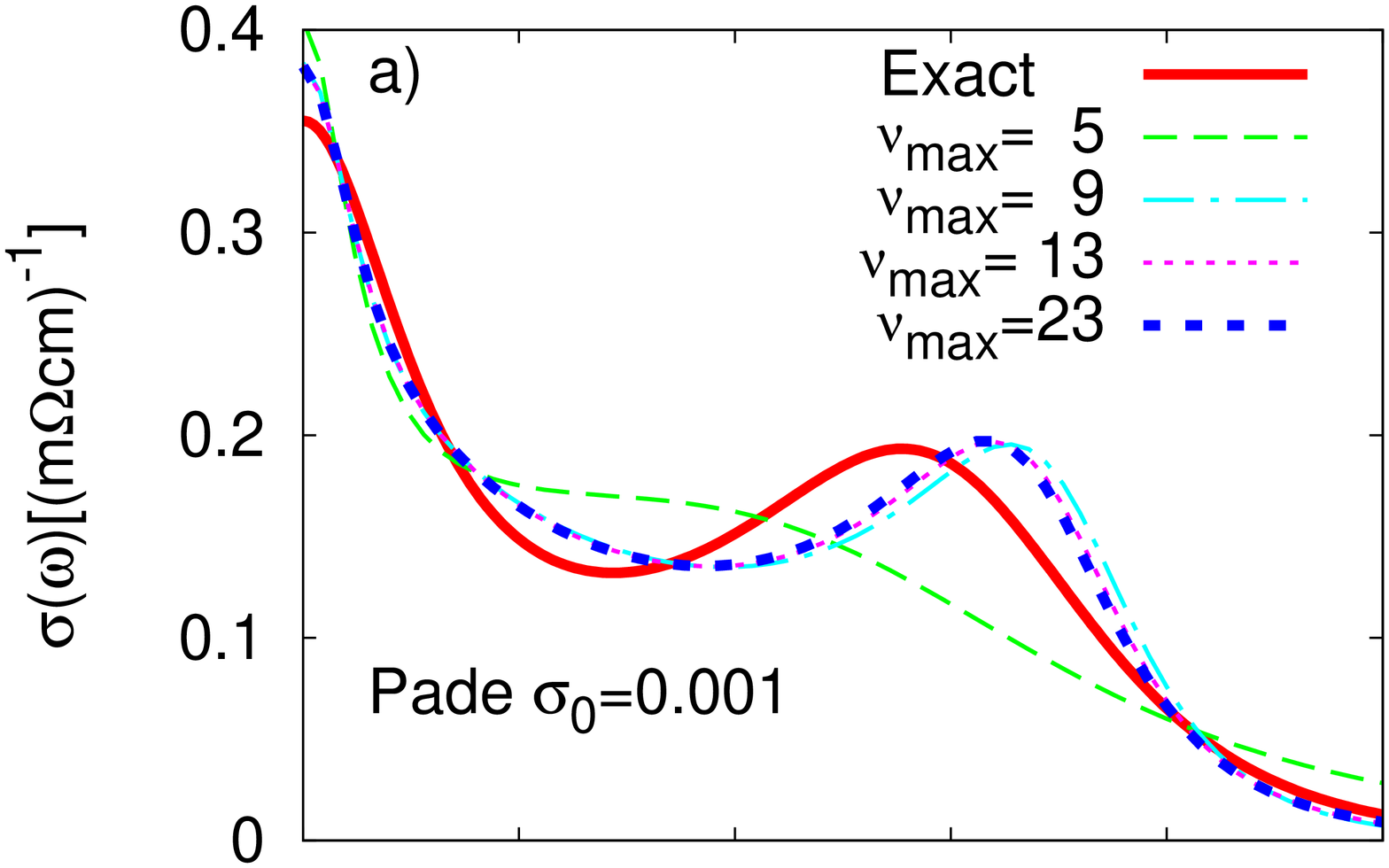}}}}
\vskip-1.34cm
{\rotatebox{-90}{\resizebox{5.7cm}{!}{\includegraphics {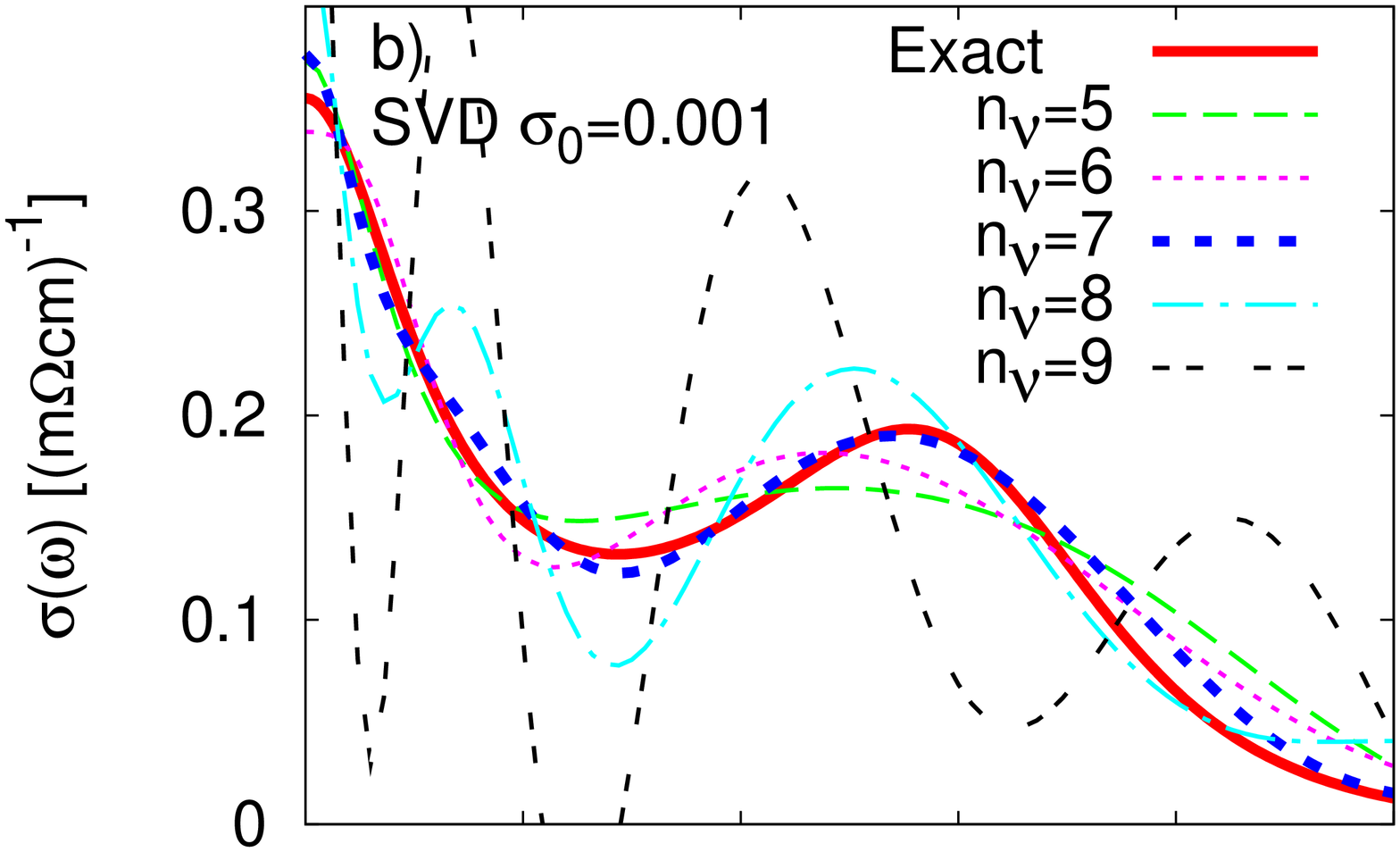}}}}
\vskip-1.34cm
{\rotatebox{-90}{\resizebox{5.7cm}{!}{\includegraphics {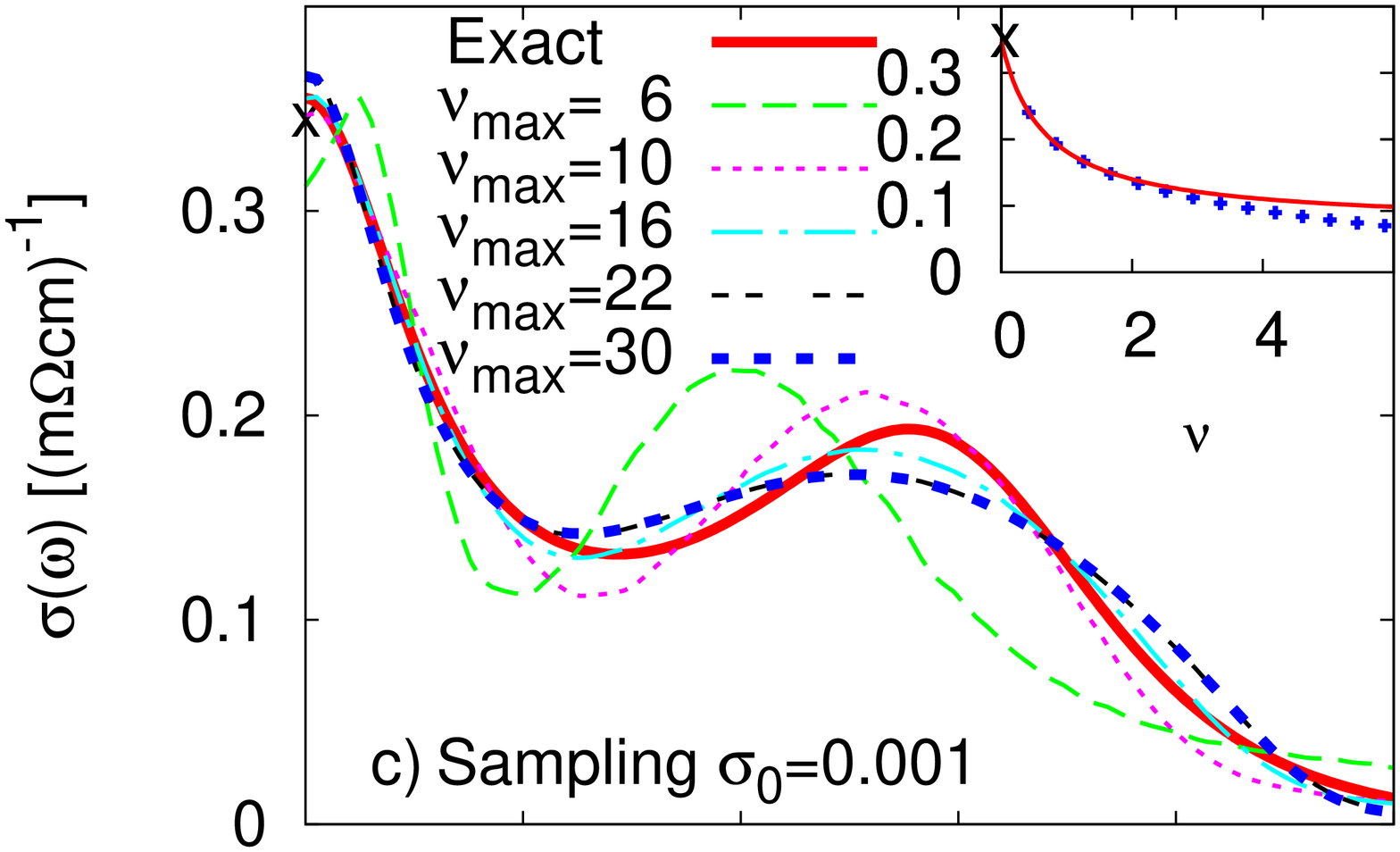}}}}
\vskip-0.78cm
{\rotatebox{-90}{\resizebox{5.7cm}{!}{\includegraphics {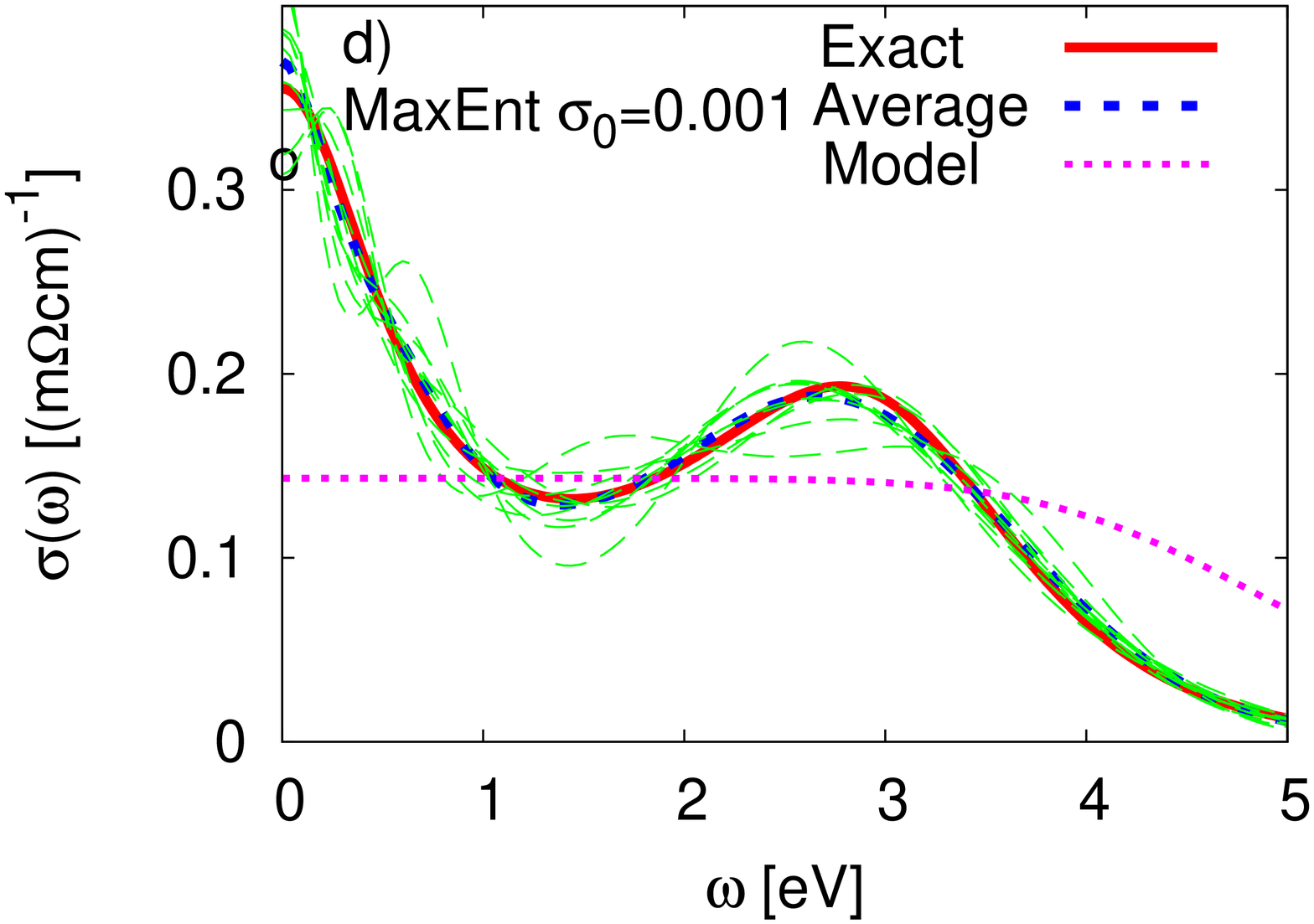}}}}
\caption{\label{fig:4}
The same as Fig. \ref{fig:2} ($\Gamma_1=0.6$) but for $\sigma_0=0.001$.
}  
\end{figure}

Fig.~\ref{fig:2}d shows the MaxEnt results. Results are shown for
each of the 10 data sets and also the average of the results is shown.
Each MaxEnt spectrum shows rather large spurious oscillations due to
the method giving too much weight to the noise. The average of these 
spectra, however, is rather good. The symbol $o$ in  Fig.~\ref{fig:2}d 
also shows the estimate in Eq.~(\ref{eq:2.7}) of $\sigma(0)$. This 
estimate is accidentally quite good, although the spectrum 
has substantial variations over the range $|\omega| \le 0.69 2\pi T$, 
and the requirement for Eq.~(\ref{eq:2.7}) is not well satisfied.
The reason is that the noise happens to make this estimate accurate, 
while in Fig.\ref{fig:4} with more accurate data the estimate is less good.

\begin{figure}
{\rotatebox{-90}{\resizebox{5.7cm}{!}{\includegraphics {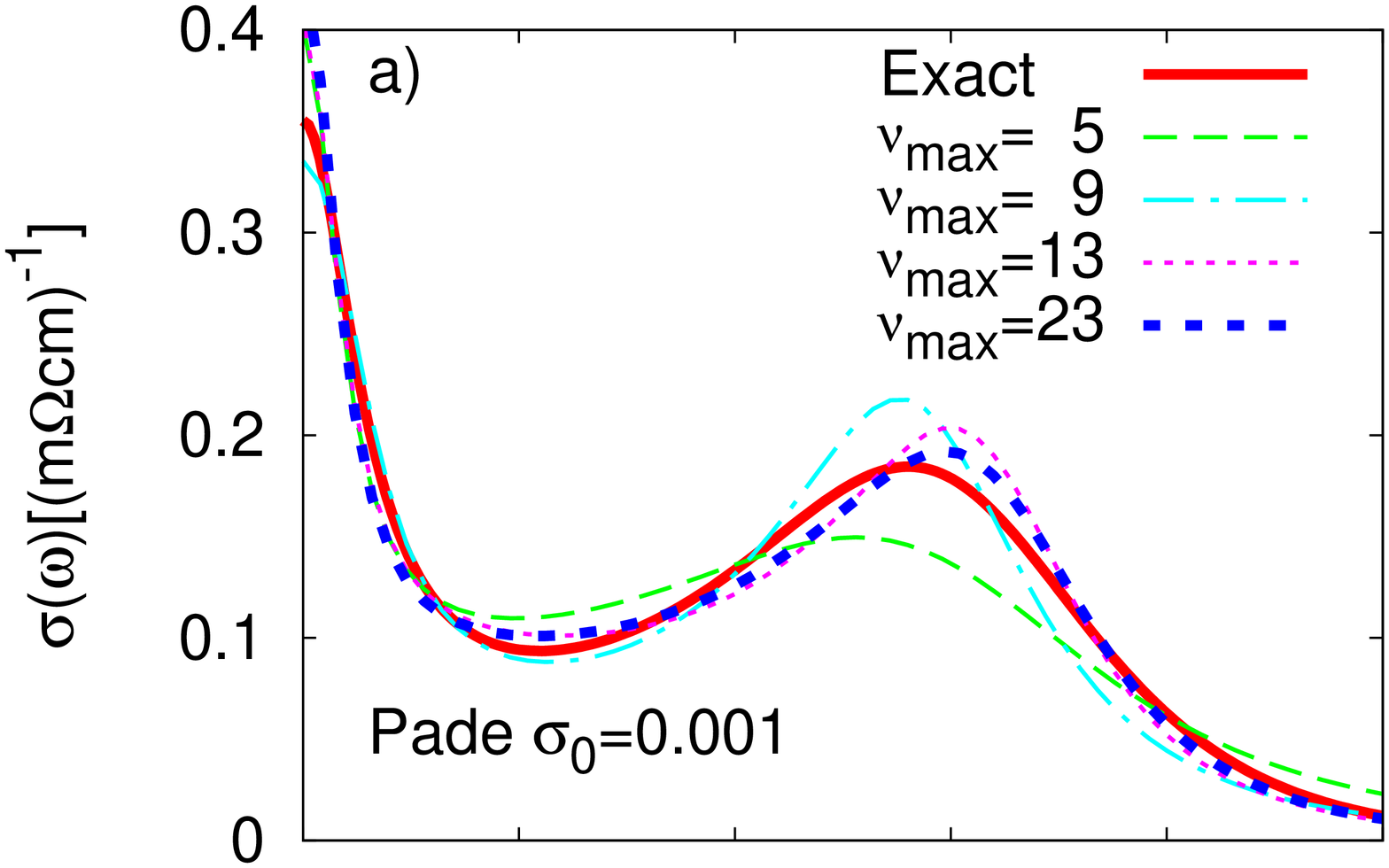}}}}
\vskip-1.34cm
{\rotatebox{-90}{\resizebox{5.7cm}{!}{\includegraphics {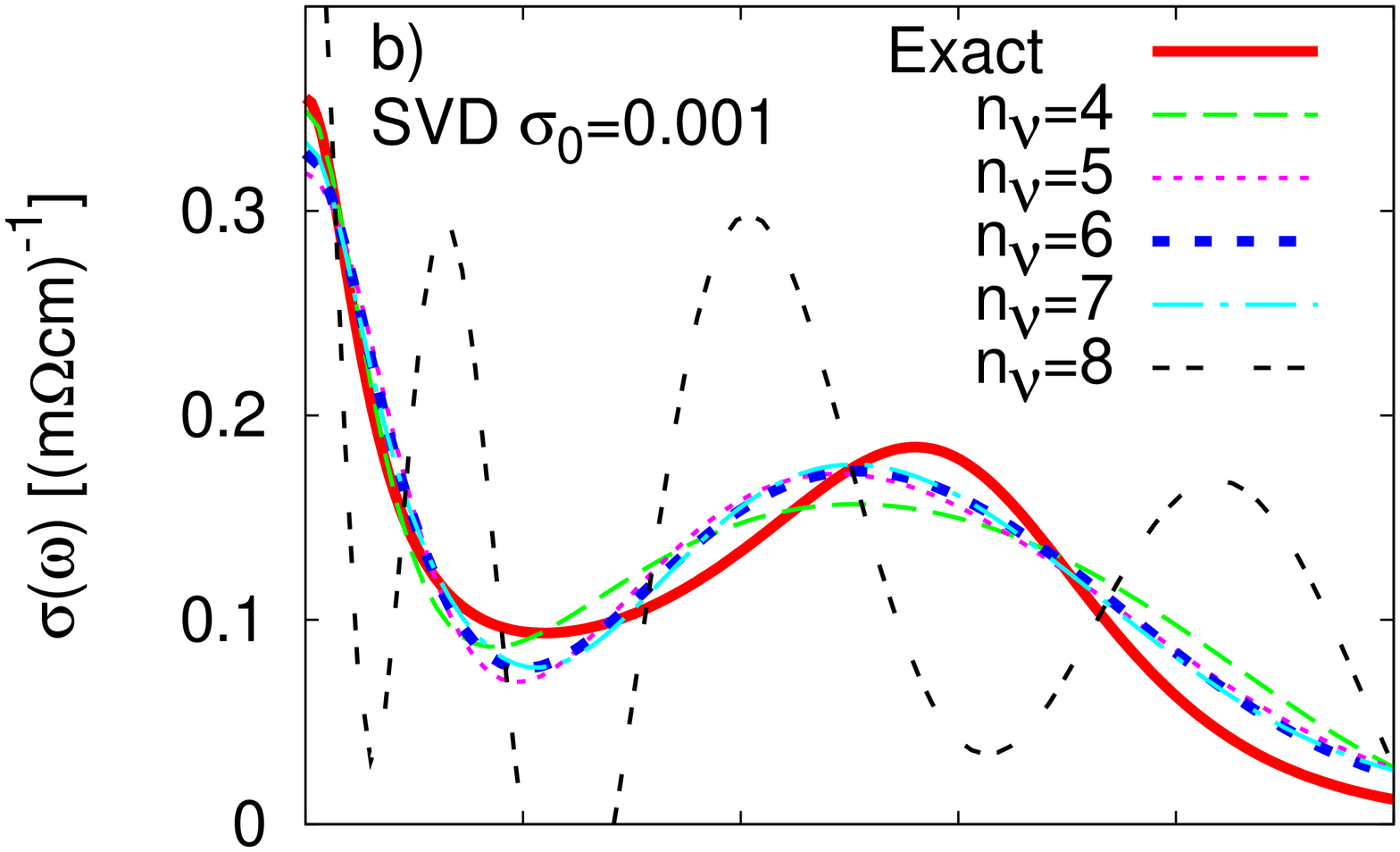}}}}
\vskip-1.34cm
{\rotatebox{-90}{\resizebox{5.7cm}{!}{\includegraphics {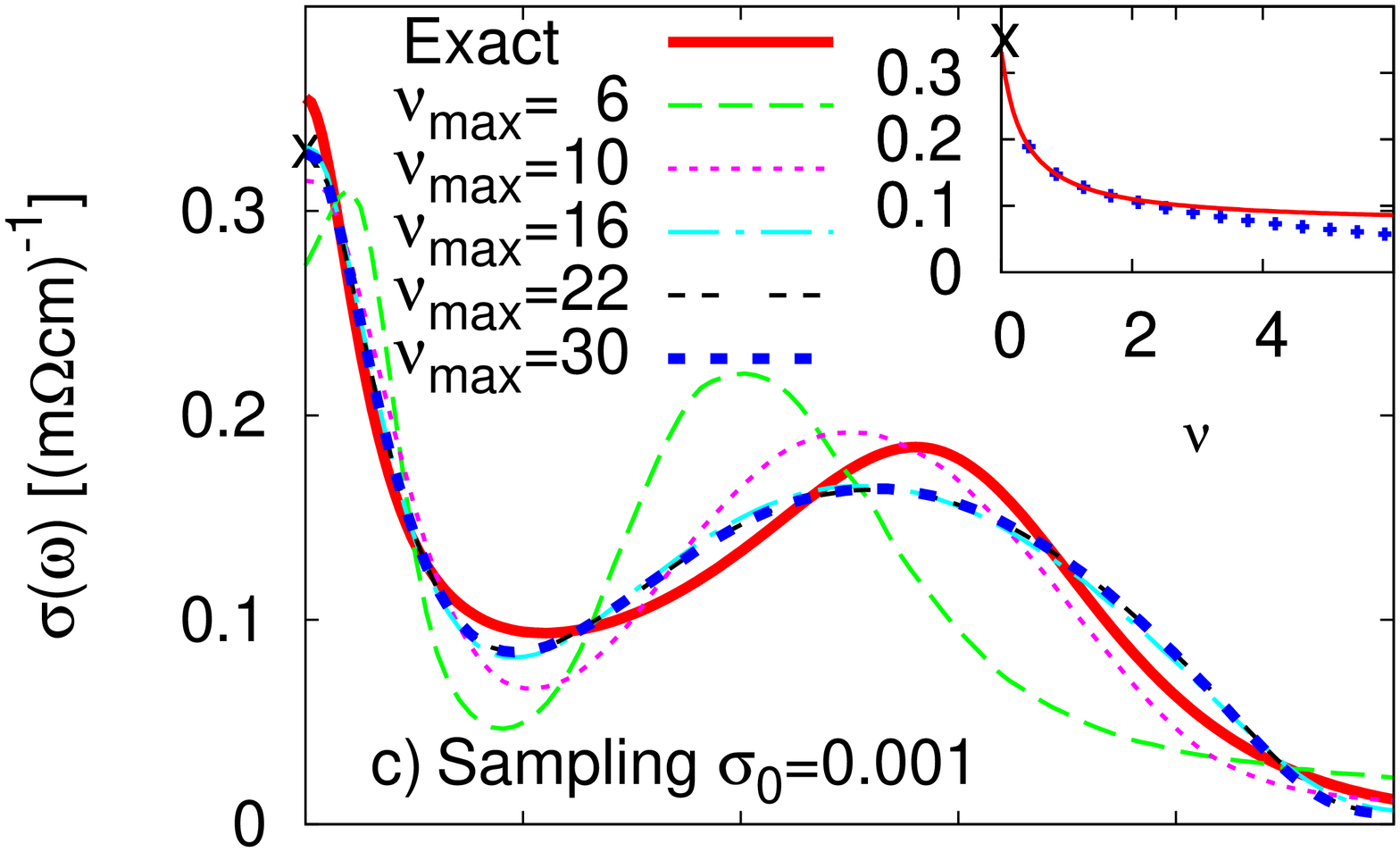}}}}
\vskip-0.78cm
{\rotatebox{-90}{\resizebox{5.70cm}{!}{\includegraphics {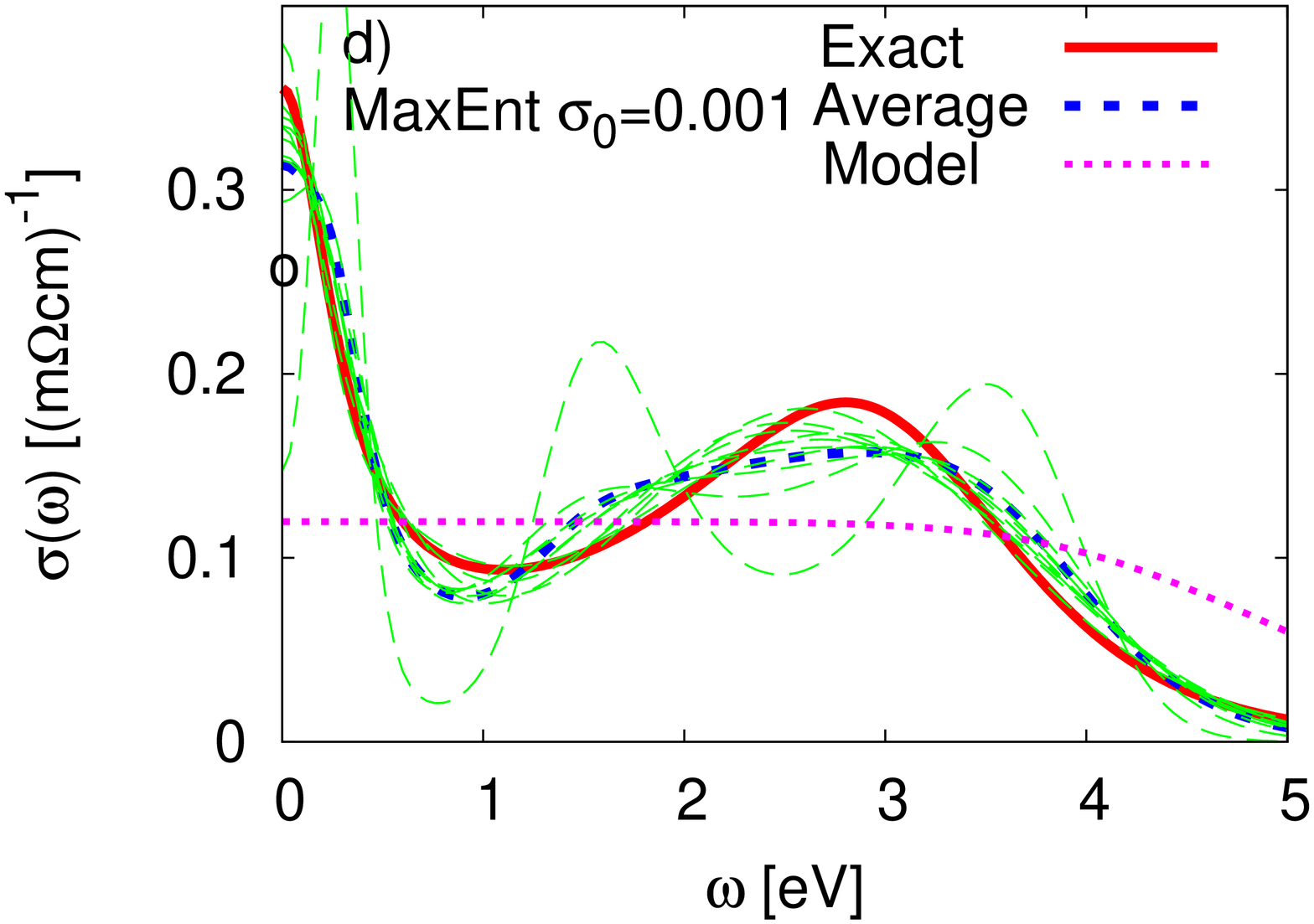}}}}
\caption{\label{fig:5}
The same as Fig. \ref{fig:2} but for $\Gamma_1=0.3$ and $\sigma_0=0.001$.
}  
\end{figure}

\begin{figure}
{\rotatebox{-90}{\resizebox{5.7cm}{!}{\includegraphics {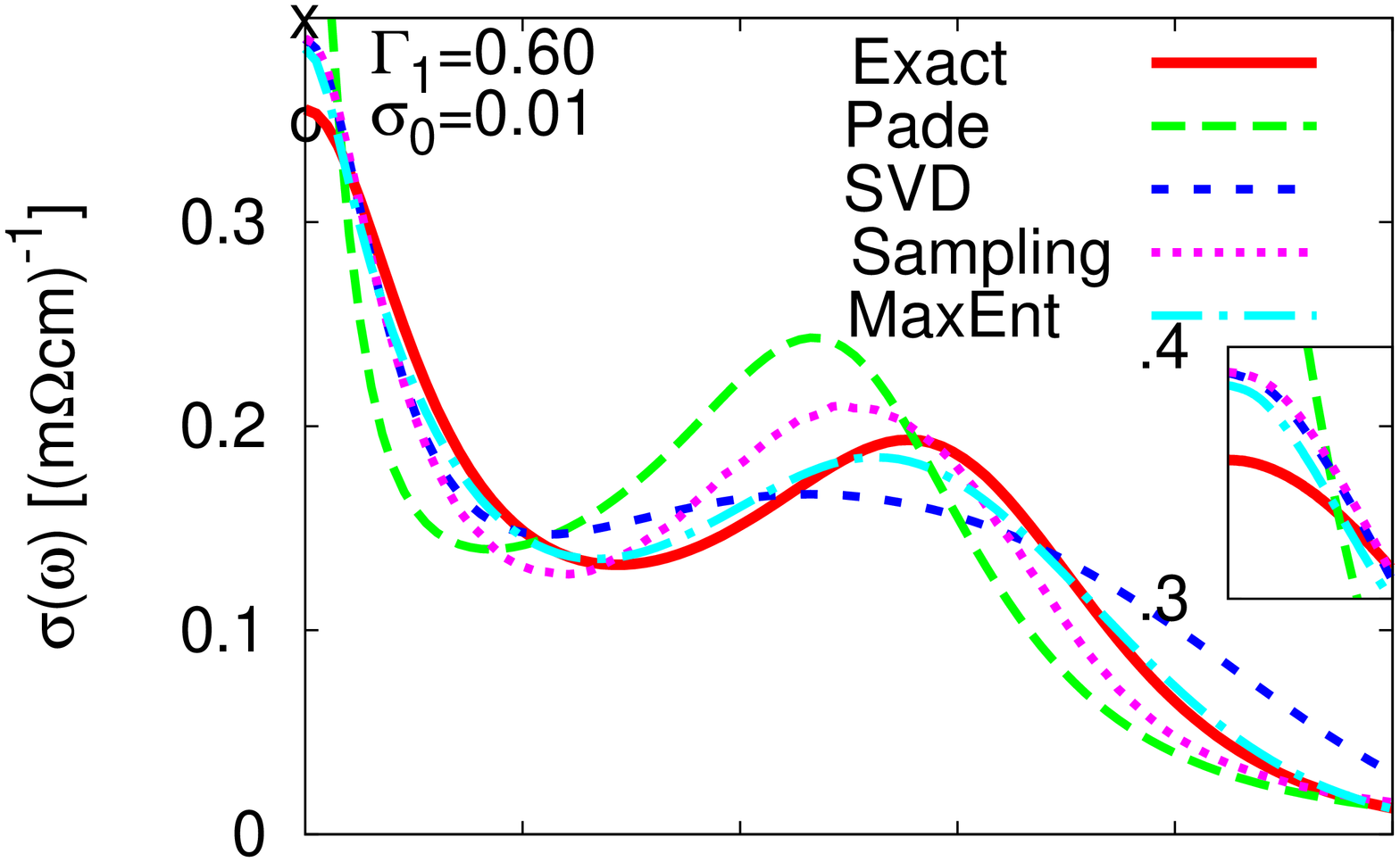}}}}
\vskip-1.34cm
{\rotatebox{-90}{\resizebox{5.7cm}{!}{\includegraphics {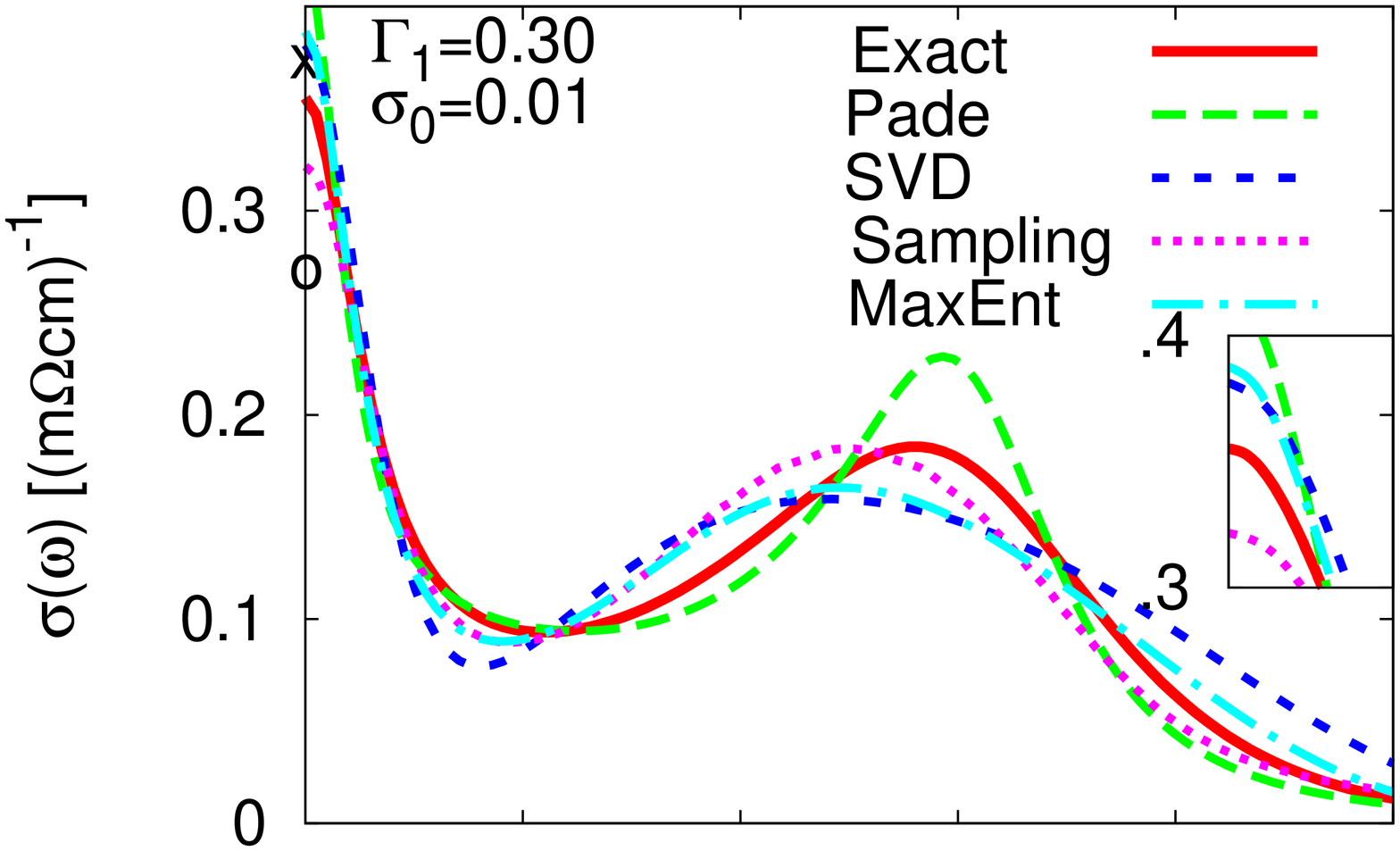}}}}
\vskip-1.34cm
{\rotatebox{-90}{\resizebox{5.7cm}{!}{\includegraphics {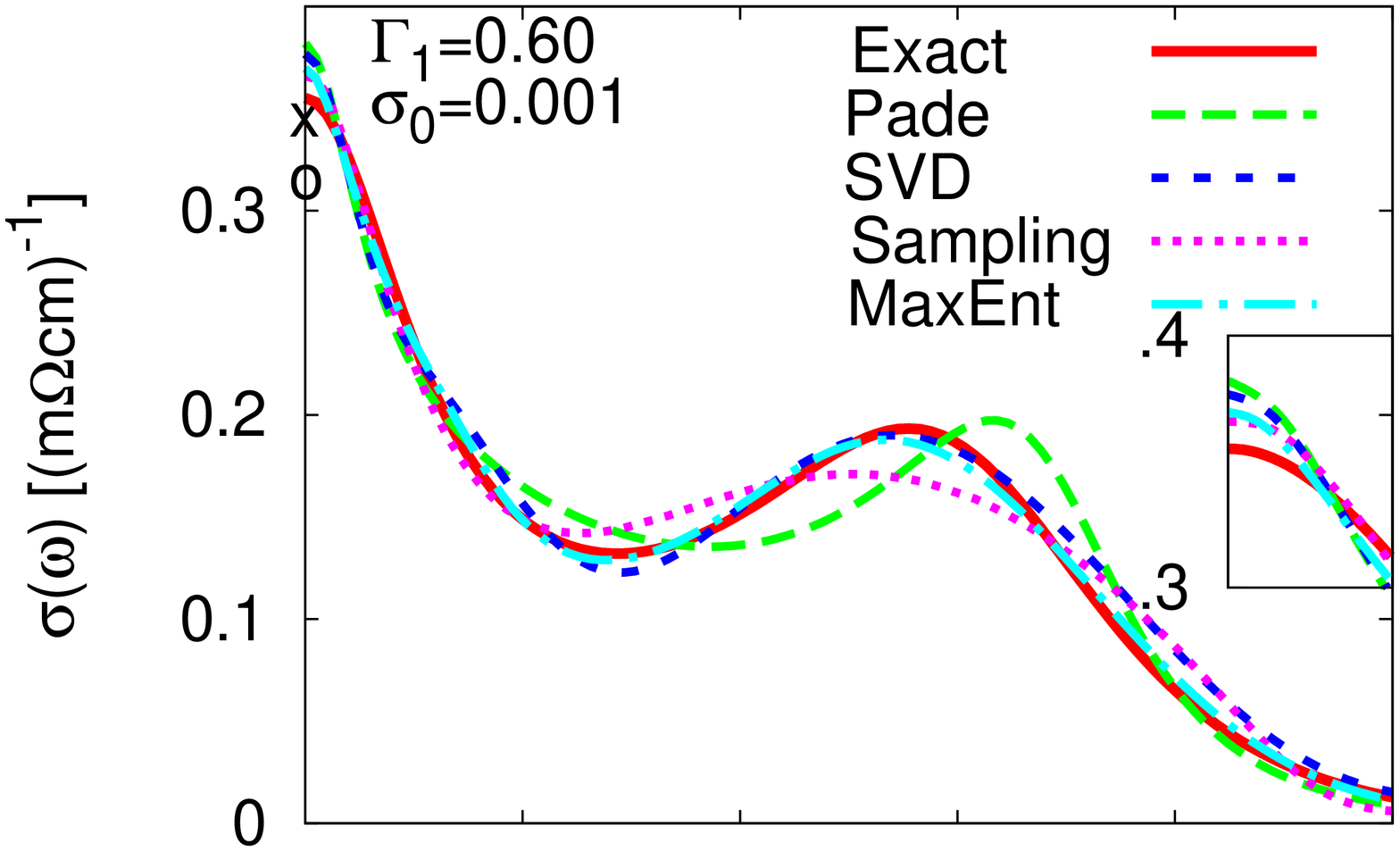}}}}
\vskip-0.78cm
{\rotatebox{-90}{\resizebox{5.7cm}{!}{\includegraphics {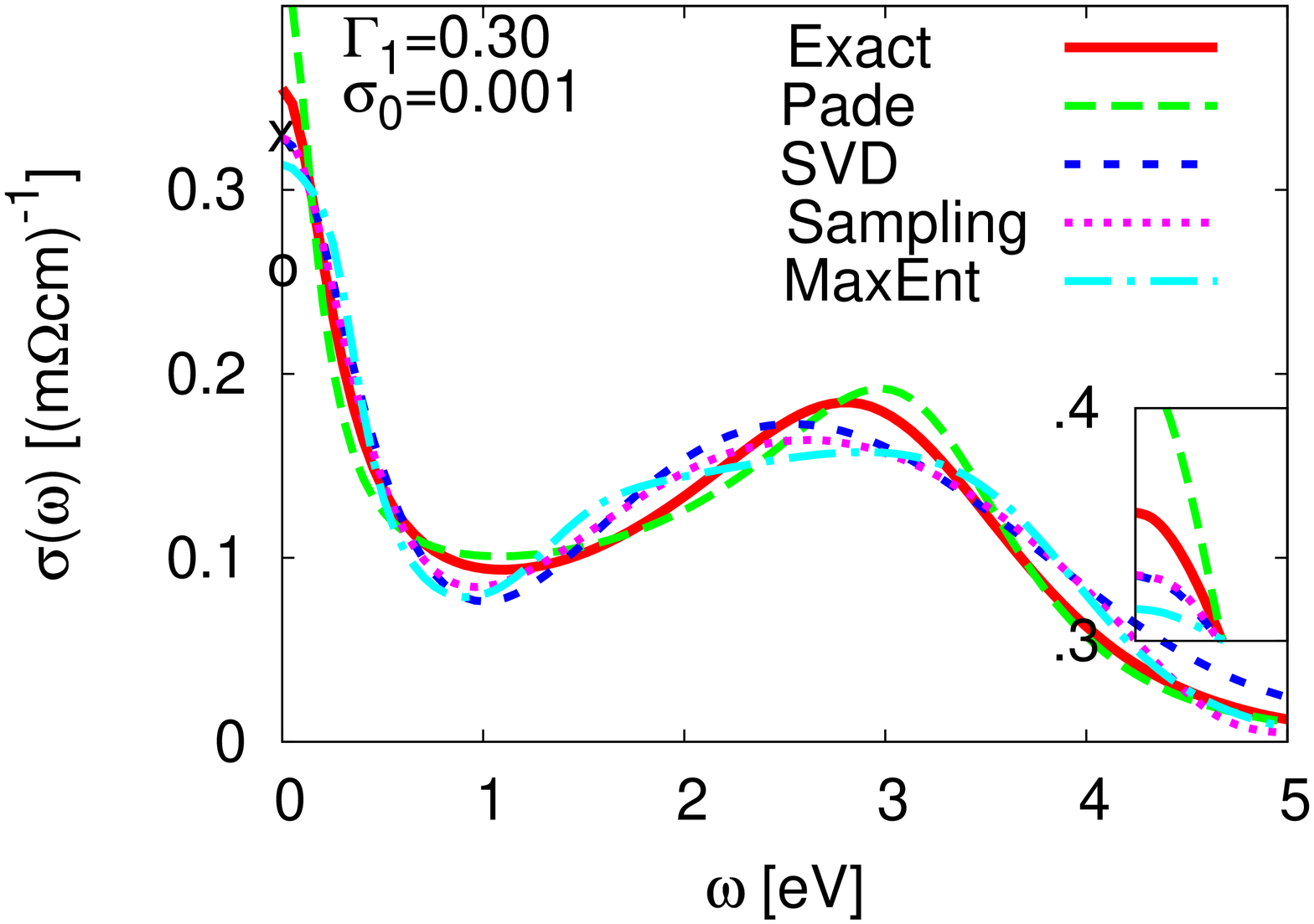}}}}
\caption{\label{fig:6}Comparison of the different methods for $\Gamma_1=$
0.6 and 0.3 and for $\sigma=$ 0.01 and 0.001. The insets show a magnified 
view in the range $0\le \omega \le0.25$.
}  
\end{figure}

\begin{figure}
{\rotatebox{-90}{\resizebox{5.7cm}{!}{\includegraphics {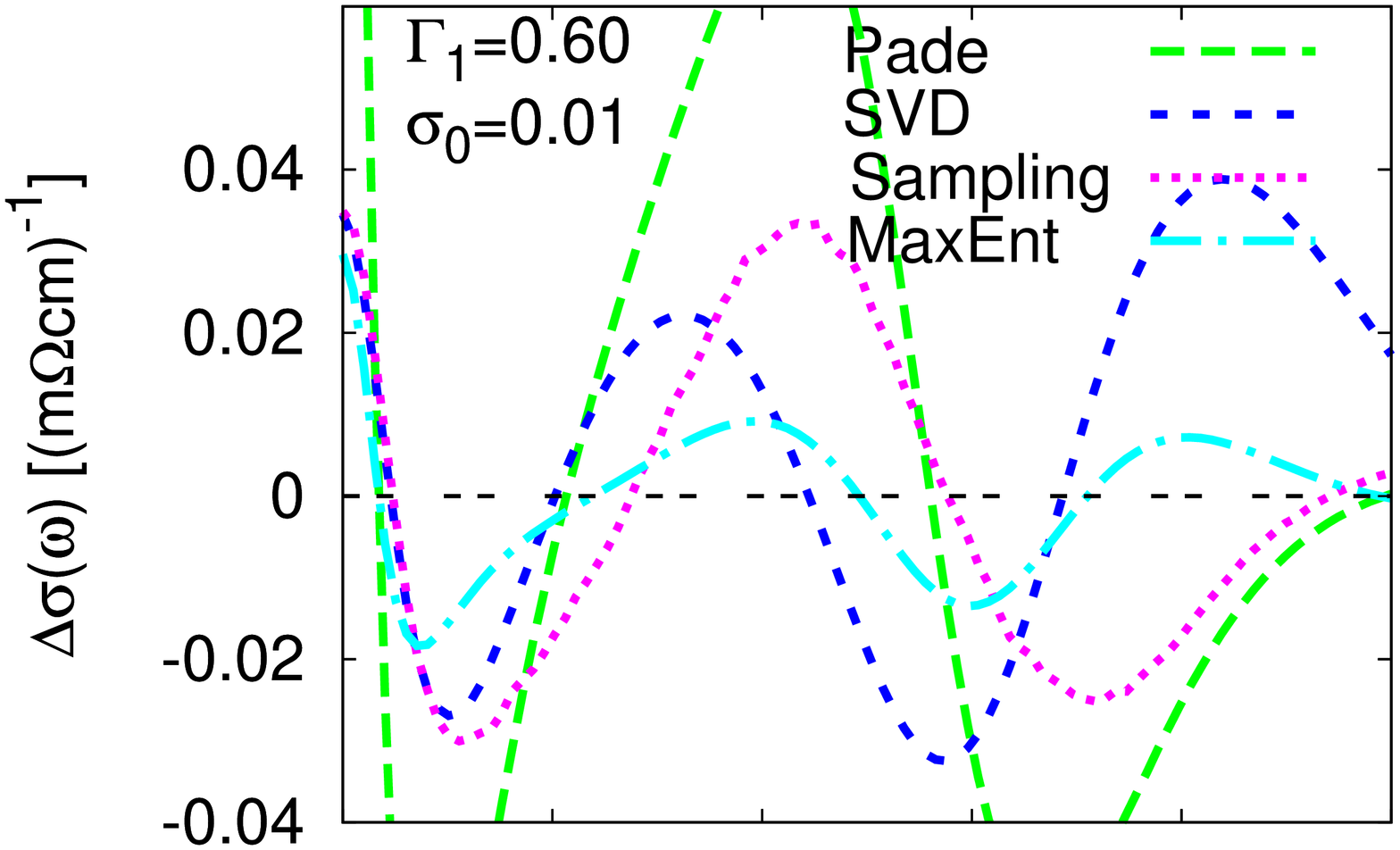}}}}
\vskip-1.34cm
{\rotatebox{-90}{\resizebox{5.7cm}{!}{\includegraphics {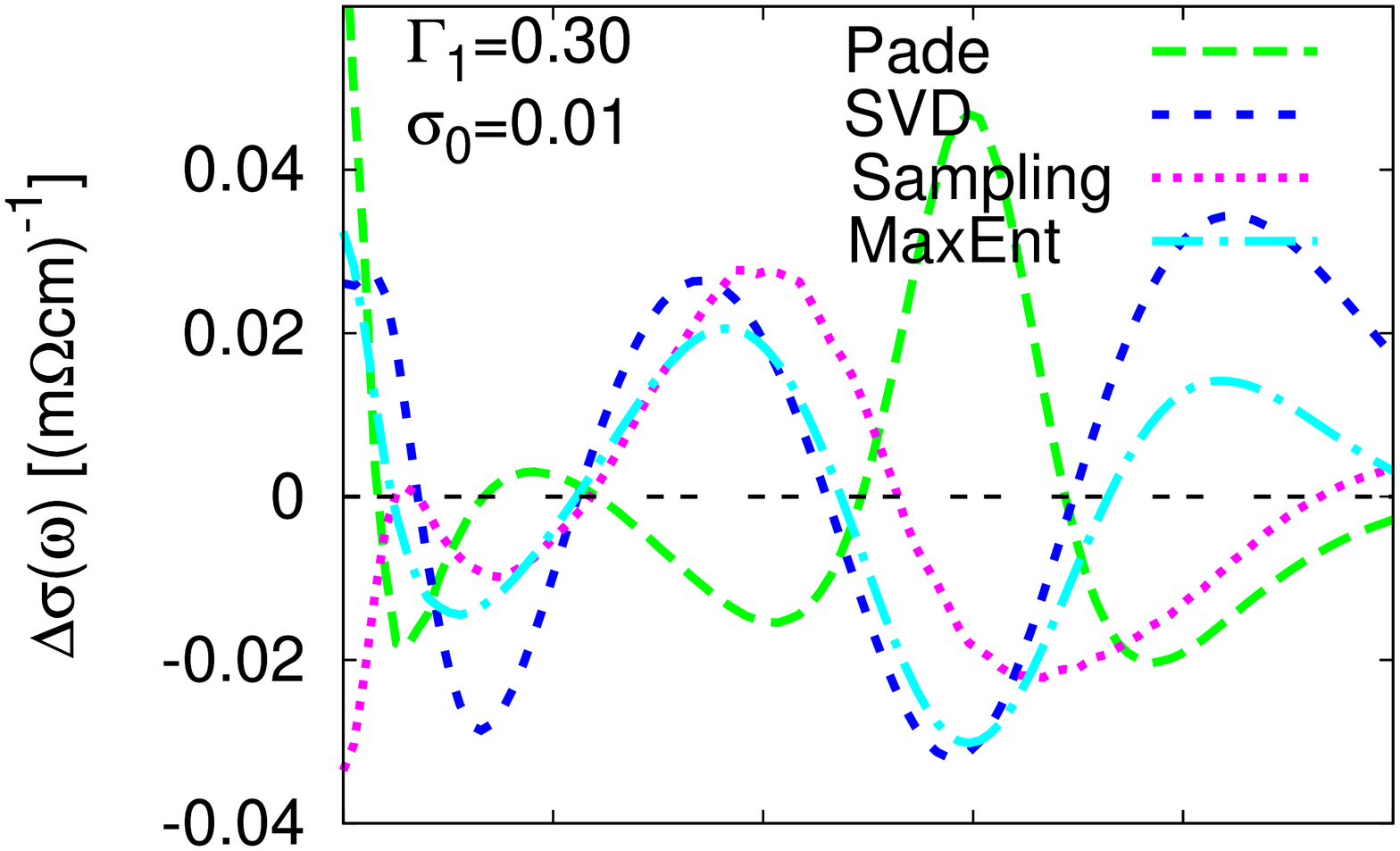}}}}
\vskip-1.34cm
{\rotatebox{-90}{\resizebox{5.7cm}{!}{\includegraphics {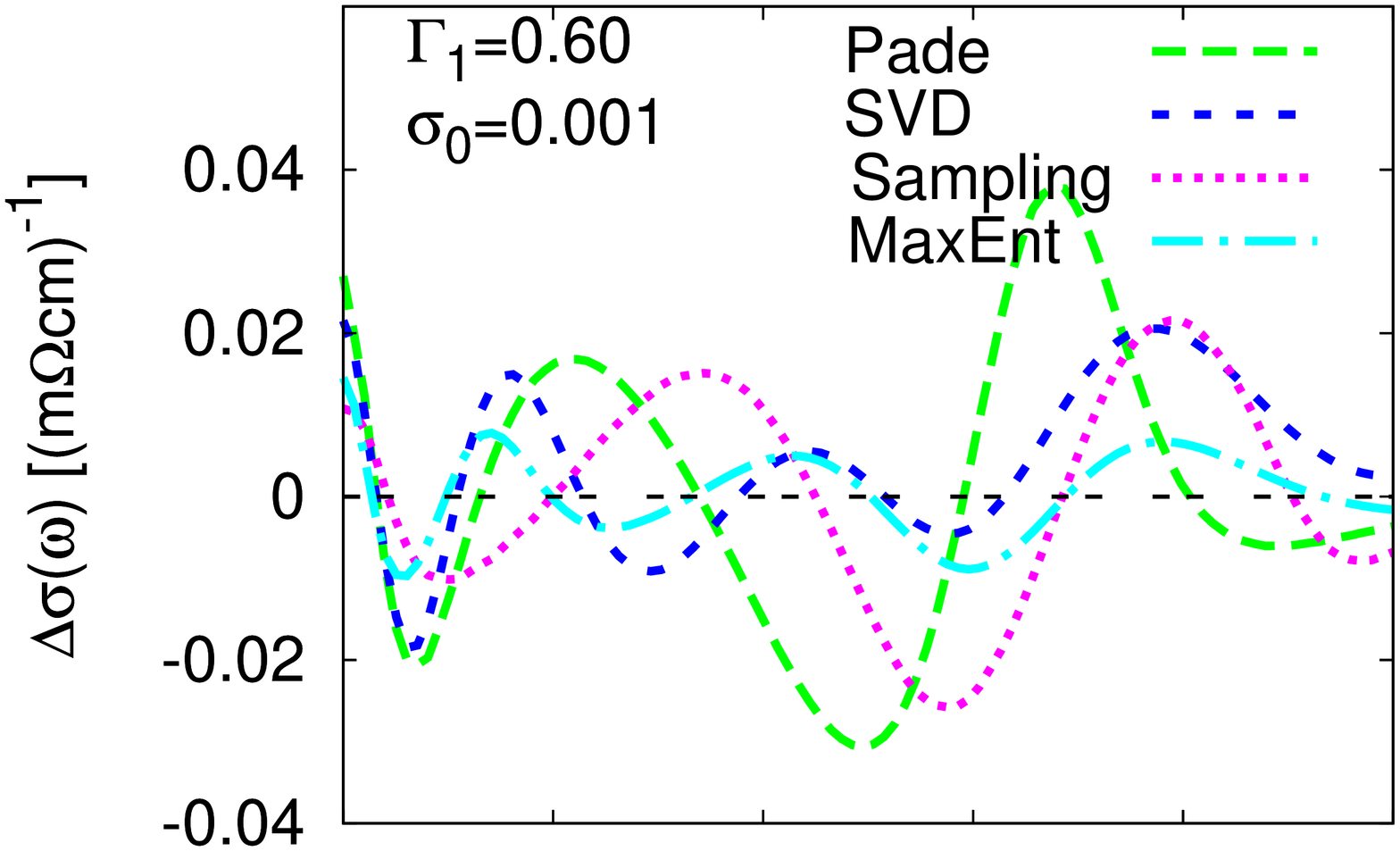}}}}
\vskip-0.78cm
{\rotatebox{-90}{\resizebox{5.7cm}{!}{\includegraphics {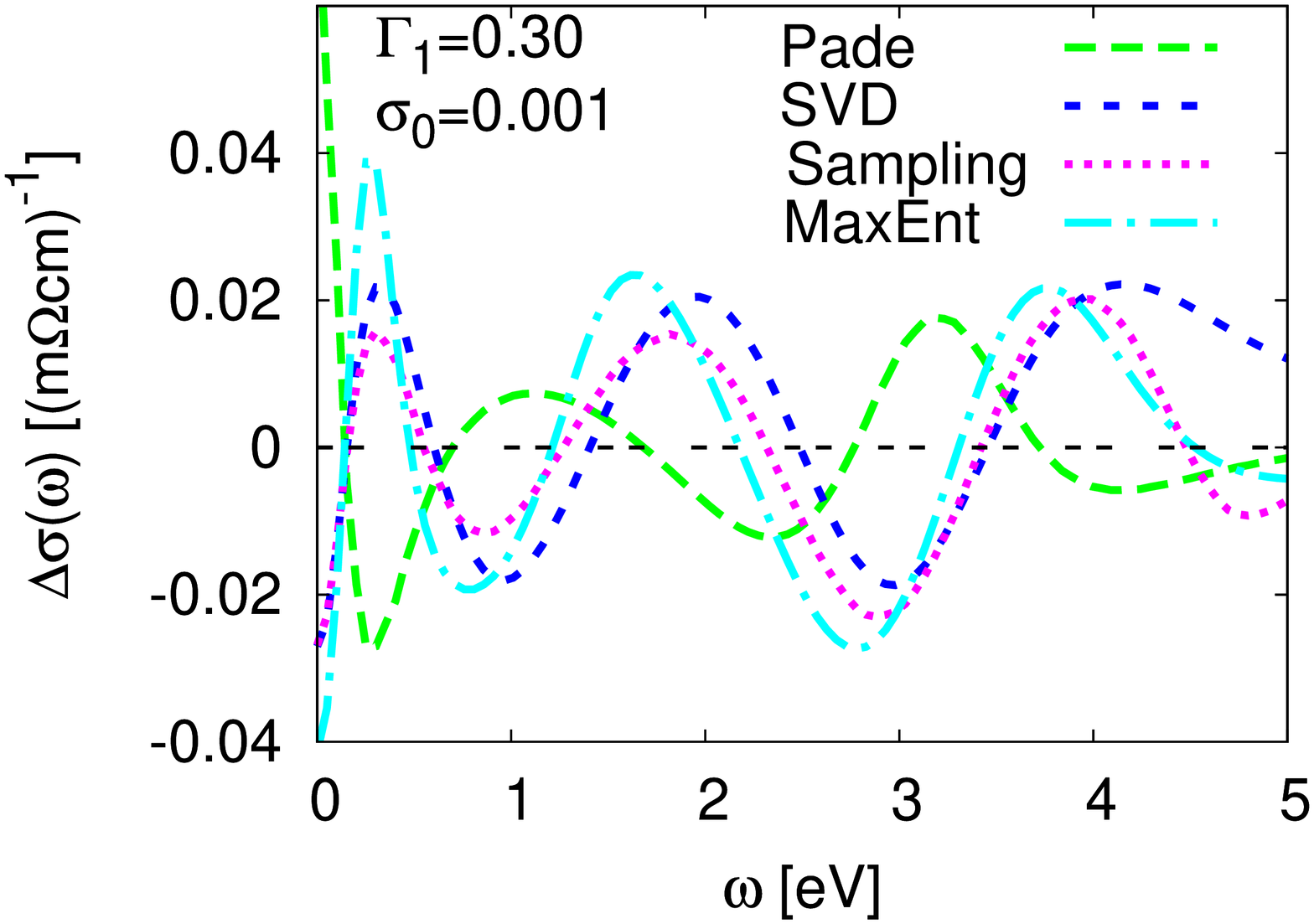}}}}
\caption{\label{fig:7}Difference between the spectrum calculated using
one of the methods and the exact spectrum for the parameters $\Gamma_1=$
0.6 and 0.3 and for $\sigma=$ 0.01 and 0.001. 
}  
\end{figure}

Fig.~\ref{fig:3} shows results for $\Gamma_1=0.3$, i.e., a narrower peak 
at $\omega=0$. The SVD, sampling and MaxEnt methods give comparable accuracy
as in Fig. \ref{fig:2}.  The accuracy of the estimates of $\sigma(0)$ from 
Eq.~(\ref{eq:2.7}) (o in Fig.~\ref{fig:3}d) is worse than 
in Fig. \ref{fig:2}, since the peak at $\omega=0$ is narrower and assumption 
behind Eq.~(\ref{eq:2.7}) is less well satisfied. The estimate from
Eq.~(\ref{eq:2.6}) ($\times$ in Fig.~\ref{fig:3}c) is of comparable
accuracy as in Fig. \ref{fig:2}. 
 
Fig.~\ref{fig:4} shows results for $\Gamma_1=0.60$, i.e., a broader peak as 
in Fig.~\ref{fig:2}, but for very  accurate data, $\sigma_0=0.001$. The 
accuracy of the Pad\'{e} approximations is now improved, as expected.
Because of the higher accuracy of the data, the optimum $n_{\nu}$ has
increased from 5 to 7 for the SVD method. This leads to an improvement compared
with Fig.~\ref{fig:2}b, although there is a small unphysical oscillation at 
$\omega\sim 1$. There is a reduced spread of the thin curves in Fig.~\ref{fig:4}d, 
representing the MaxEnt result for each individual data set. The average is 
only marginally improved.                                                                 
Fig.~\ref{fig:5} shows high accuracy data for a narrow peak, $\Gamma_1=0.3$. 

In Fig.~\ref{fig:6} we compare the different methods for the two different
spectra ($\Gamma$ = 0.3 and 0.6) and for the two accuracies ($\sigma_0$ =0.01 and 0.001)
considered here. Since the value of $\sigma(0)$ is of particular interest, we show results for
small $\omega$ ($\le 0.25$) in the insets. Typically the SVD, sampling and MaxEnt methods
are of comparable accuracy, while the Pade approximation tends to overestimate
 $\sigma(0)$. The differences between the results of these methods and the exact result
are shown in Fig.\ref{fig:7}. 

\section{Correlation in imaginary time}
We have so far generated data for imaginary frequencies and then added Gaussian
noise. The noise for different frequencies is uncorrelated and the covariant matrix
\begin{eqnarray}\label{eq:7.1}
&& C_{ik}={1\over M(M-1)}    \\
&& \times \sum_{j=1}^M [\bar \Pi(\nu_i)-\Pi^{(j)}(\nu_i)][\bar \Pi(\nu_k)-\Pi^{(j)}(\nu_k)] \nonumber
\end{eqnarray}
is approximately diagonal. Here $\bar \Pi(\nu_i)$ is the average over the $M$ samples 
$\Pi^{(j)}(\nu_i)$. If the data are obtained from a QMC calculation, $C$ is in general
not diagonal. There is then a need to make a transformation to a diagonal covariant
matrix. Here we follow Jarrell and Gubernatis.\cite{Jarrell} A matrix $U$ is
found such that
\begin{equation}\label{eq:7.2}
C^{'}=U^{-1}CU
\end{equation}
is diagonal. The data and kernel are then transformed to the new representation
\begin{equation}\label{eq:7.3}
K^{'}=U^{-1}K \hskip1cm \Pi^{'}=U^{-1}\Pi
\end{equation}
and the diagonal elements of $C^{'}$ are used to define a new likelihood function.
The result is that some of the diagonal elements of the covariant matrix are
now larger, implying less accurate data than one might have thought. This does 
not, however, change the general conclusions above.

\section{Conclusions}\label{sec:6}

We have compared different methods for analytically continuation of imaginary
axis data to real frequencies for the optical conductivity. We transform
spectra from the real frequency axis to the imaginary axis and add statistical
noise. These data are then transformed back to the real axis using the  different
analytical continuation methods. By comparing with the original spectrum, 
we can compare the accuracy of these methods. Typically, these methods 
have problems if the spectra have features on a much smaller energy range 
than $2\pi T$. Due to the thermal broadening of  physical spectra, this may 
not be a serious problem in many cases. Here we have focused on two cases where 
the relevant energy scale, $\Gamma_1$ is 0.3 or 0.6 compared with $2\pi T=0.42$.

 We also considered two methods for obtaining $\sigma(0)$ directly, 
Eq.~(\ref{eq:2.7}) and extrapolation of $\gamma(\nu)$ 
in Eq.~(\ref{eq:2.6}) to $\nu=0$. The method based on Eq.~(\ref{eq:2.7}) 
tends to underestimate $\sigma(0)$, in particular if $\sigma(\omega)$ has 
a narrow Drude peak, while the extrapolation of Eq.~(\ref{eq:2.6}) is
typically more accurate.
 
Calculations for the cases considered in this paper as well as for results
from DCA typically gives larger values for $\sigma(0)$ in the Pad\'{e} approximation
than from the SVD, sampling and MaxEnt approaches. The Pad\'{e} approximation
generally tends to give somewhat less accurate results than the other three
methods. Sometimes unphysical results are obtained due to poles close to the real
axis. The other three methods tend to give results of comparable accuracy.
We nevertheless find it very useful to use all three methods. This provides
cross checks and gives a somewhat better idea about what the true spectrum 
may look like.

\section{Acknowledgments}
We would like to thank M. Jarrell for making his MaxEnt program available
and C. Creffield for providing his SVD program.
One of us (GS) acknowledges support from the FWF ``Lise-Meitner'' grant n. M1136.


\begin{thebibliography}{99}

\bibitem{Scalapino}R. Blankenbecler, D.J. Scalapino, R.L. Sugar,
Phys. Rev. D {\bf 24}, 2278 (1981).

\bibitem{DCA}T. Maier, M. Jarrell, T. Pruschke and M. H. Hettler, Rev.
Mod. Phys. {\bf  77}, 1027 (2005).

\bibitem{RubtsovCT}A.N. Rubtsov and A.I. Lichtenstein, JETP Letters
{\bf 80}, 61 (2004), P. Werner and A.J. Millis, Phys. Rev. B.
{\bf 74}, 155107 (2006).


\bibitem{Jarrell}M. Jarrell and J.E. Gubernatis, Phys. Rep.
{\bf 269}, 133 (1996).

\bibitem{MEMref}R. N. Silver, D. S. Sivia, and J. E. Gubernatis, Phys. Rev. B
{\bf 41}, 2380 (1990); J. E. Gubernatis, M. Jarrell, R. N. Silver, and
D. S. Sivia, Phys. Rev. {\bf 44}, 6011 91991); W. von der Linden,
Appl. Phys. A {\bf 60}, 155 (1995).

\bibitem{SVD}M. Bertero, C. De Mol, and E.R Pike, Inverse Problems {\bf 1},
301 (1985); M. Bertero and E.R Pike, {\it Handbook of Statistics}, edited by 
N.K. Bose and C.R. Rao (Elsevier Sience Publishers, New York, 1993) Vol. 10;
M. Bertero, P. Branzi, E.R. Pike, and L. Rebolia, Proc. R. Soc. London 
A {\bf 415}, 257 (1988).

\bibitem{creffield}C.E. Creffield, E.G. Klepfish, E.R. Pike, and S. Sarkar,
Phys. Rev. Lett. {\bf 75}, 517 (1995).

\bibitem{Rubtsov}I.S. Krivenko and A.N. Rubtsov, arXiv:cond-mat/0612233

\bibitem{Vidberg}H. J. Vidberg and J. W. Serene, J. Low Temp. Phys.
{\bf 29}, 179 (1977).

\bibitem{Baker}G. A. Baker, Jr., {\it Essentials of Pad$\acute e$ 
approximants} Academic, New Yok, 1975, p 100ff.


\bibitem{Mishchenko}A. S. Mishchenko, N. V. Prokof'ev, A. Sakamoto,
and B. V. Svistunov, Phys. Rev. B {\bf 62}, 6317 (2000).

\bibitem{Kiamars}K. Vafayi and O. Gunnarsson, Phys. Rev. B {\bf 76}, 035115 (2007).


\bibitem{Bayes}R. T. Cox, {\it The Algebra of Probable Inference},
John Hopkins University Press, Baltimore, Maryland (1961);
D. S. Sivia, {\it Data Analysis - A Bayesian tutorial}, Clarendon Press,
Oxford (1996).

\bibitem{Hanke}S. Hochkeppel, F. F. Assaad, and W. Hanke, Phys. Rev. B 
{\bf 77}, 205103 (2008).

\bibitem{Ceperley}F. Lin, M. A. Morales, K. T. Delaney, C. Pierleoni, 
R. M. Martin, and D. M. Ceperley, Phys. Rev. Lett. {\bf 103}, 256401 
(2009).

\bibitem{Skilling}See, e.g., J. Skilling, J. Microscopy {\bf 190},
28 (1997).

\bibitem{White}S.R. White, Phys. Rev. B 44, 4670 (1991).  

\bibitem{batching}O. Gunnarsson, M. W. Haverkort, and G. Sangiovanni,
Phys. Rev. B {\bf 81}, 155107 (2010).  

\bibitem{Metropolis}N. Metropolis, A. W. Rosenbluth,M. N. Rosenbluth,
A. H. Teller, and E. Teller, J. Chem. Phys. {\bf 21}, 1087 (1953).




\end{thebibliography}
\end{document}